\documentclass[traditabstract]{aa} 

\usepackage{graphicx}
\usepackage{txfonts}
\usepackage{bigstrut}
\begin{document}

   \title{Synthetic photometry for carbon-rich giants}

   \subtitle{III. Tracing the sequence of mass-losing galactic C-type Miras\thanks{Appendices are only available in electronic form at
\newline 
\texttt{http://www.aanda.org}}}

   \author{W.~Nowotny\inst{1}   \and
           B.~Aringer\inst{1,2}   \and
           S.~H\"ofner\inst{3}  \and
           K.~Eriksson\inst{3}
           }

   \institute{University of Vienna, Department of Astrophysics, 
              T\"urkenschanzstra{\ss}e 17, A-1180 Wien, Austria\\
              \email{walter.nowotny@univie.ac.at}
         \and
              INAF -- Padova Astronomical Observatory, 
              Vicolo dell'Osservatorio 5, 35122 Padova, Italy
         \and
              Department of Physics and Astronomy, 
              Division of Astronomy and Space Physics,
              Uppsala University,
              Box 516, SE-75120 Uppsala, Sweden\\
             }

   \date{Received; accepted}

\titlerunning{Synthetic photometry for carbon-rich giants III.}
\authorrunning{W. Nowotny et al. }

 
\abstract{Late-type giant stars in the evolutionary stage of the asymptotic giant branch increasingly lose mass via comparatively slow but dense stellar winds. Not only do these evolved red giants contribute in this way to the enrichment of the surrounding interstellar medium, but the outflows also have a substantial influence on the spectro-photometric appearance of such objects. In the case of carbon-rich atmospheric chemistries, the developing cool circumstellar envelopes contain dust grains mainly composed of amorphous carbon. With increasing mass-loss rates, this leads to more and more pronounced circumstellar reddening. With the help of model calculations we aim at reproducing the observational photometric findings for a large sample of well-characterised galactic C-type Mira variables losing mass at different rates. We used dynamic model atmospheres, describing the outer layers of C-rich Miras, which are severly affected by dynamic effects. Based on the resulting structures and under the assumptions of chemical equilibrium as well as LTE, we computed synthetic spectra and synthetic broad-band photometry (Johnson-Cousins-Glass BVRIJHKL$^\prime$M). A set of five representative models with different stellar parameters describes a sequence from less to more evolved objects with steadily increasing mass-loss rates. This allowed us to study the significant influence of circumstellar dust on the spectral energy distributions and the (amplitudes of) lightcurves in different filters. We tested the photometric properties (mean NIR magnitudes, colours, and amplitudes)  and other characteristics of the models (mass-loss rates, periods, and bolometric corrections) by comparing these with the corresponding observational data adopted from the literature. Using different kinds of diagrams we illustrate where the models are located in a supposed evolutionary sequence defined by observed C-type Mira samples. Based on comparisons of galactic targets with empirical relations derived for C stars in the Large Magellanic Cloud we discuss the relevance of metallicity and excess carbon \mbox{(C--O)} for the development of dust-driven winds. Having investigated the dynamic model atmospheres from different (mainly photometric) perspectives, we conclude that our modelling approach (meaning the combination of numerical method and a suitable choice of model parameters) is able to describe C-rich long-period variables over a wide range of mass-loss rates, i.e., from moderately pulsating objects without any dusty wind to highly dust-enshrouded Carbon Miras. Thus, we can trace the observed sequence of C-type Miras, which is mainly determined by the mass loss.}

   \keywords{stars: AGB and post-AGB --
             stars: atmospheres --
             stars: carbon --
             stars: variables: general --
             stars: mass-loss --
             circumstellar matter
            }

   \maketitle

\section{Introduction}
\label{s:intro}

This work is the third in a series of publications dealing with synthetic spectra and photometry of carbon-rich late-type giants. In the first part, Aringer et al. (\cite{AGNML09}; below Paper\,I) presented a grid of hydrostatic, dust-free model atmospheres and the corresponding photometric results. It was found that the synthetic photometry fits observational data reasonably well for objects with effective temperatures above $\approx$\,2800\,K. 

In the second part, Nowotny et al. (\cite{NoAHL11}; from now on Paper\,II) extended this study to more evolved C-rich stars for which dynamic effects become important. With outer atmospheric layers strongly influenced by (i) shock waves caused by pulsations of the stellar interiors and (ii) the development of dusty stellar winds, a different modelling approach was needed when trying to reproduce the observational properties of such objects. In Paper\,II we presented one selected dynamic model atmosphere that is representative of a typical C-type Mira variable losing mass at an intermediate rate and that accounts for these two dynamic effects. Simulating the pulsation with a variable inner boundary condition results in large-amplitude photometric variations similar to observed lightcurves. The circumstellar dust composed of amorphous carbon (amC) grains leads to even higher amplitudes, as well as to redder and more realistic colours compared with the hydrostatic case. Both findings are characteristic of long-period variables (LPVs), which are evolved red giant stars in the evolutionary phase of the asymptotic giant branch (AGB). We investigated in detail the resulting synthetic lightcurves and variations in different colour indices as a function of phase $\phi_{\rm bol}$. The comparison proved that these modelling results resemble the observed light variations of the C-rich Mira RU\,Vir reasonably well.

The choice of model in Paper\,II, i.e. one with an intermediate mass-loss rate, was motivated by two reasons: (i) to study the effects of pulsation and dust in a typical Mira, and (ii) to show that our modelling approach gives realistic results for stars where the emergent flux is a mixture of photospheric and circumstellar contributions. Extreme cases, i.e. dust-free, only mildly pulsating stars on the one hand, or objects with optically thick dusty envelopes on the other, can be reproduced rather well with hydrostatic atmosphere models (Paper\,I) or earlier generations of dust-driven wind models (e.g. H\"ofner \& Dorfi \cite{HoefD97}, Winters et al. \cite{WLJHS00}). However, to achieve a realistic representation of both the stellar atmosphere and the dusty outflow, it is necessary to apply models that combine time-dependent dynamics, detailed non-equilibrium dust formation, and frequency-dependent radiative transfer (H\"ofner\ et al. \cite{HoGAJ03}, Nowotny et al. \cite{NowHA10}, Paper\,II).

In the present paper we study the influence of stellar parameters on dust-driven outflows and the resulting photometric properties, illustrated by a sequence of models representing different phases of AGB evolution. We demonstrate that our modelling approach allows us to cover a wide range of observed properties by adjusting a few well-defined input parameters. In this way we can compute realistic synthetic photometry for models ranging from dust-free pulsating atmospheres through objects with intermediate mass-loss rates to stars with optically thick dusty outflows. 

For clarity, it should be mentioned here that the emergent wind properties (dust characteristics, mass-loss rates, velocities, optical depths, etc.) are results -- not additional input parameters -- of our models that link stellar atmospheres and circumstellar envelopes in a self-consistent way. In contrast to other often-used approaches (e.g. DUSTY code) where the outcome of a dust radiative transfer calculation is controlled by the choice of input parameters (i.e. properties of the circumstellar dust shells) quite directly, the relation between basic stellar properties and synthetic observables is non-trivial in our \textit{ab initio} modelling. The non-linearities inherent to the self-consistent dynamical modelling make a direct fitting of specific objects by fine-tuning of input parameters complicated. Notably, it is not possible to create arbitrary combinations of atmospheres and circumstellar envelopes, and certain sets of stellar parameters will not lead to outflows at all. On the positive side, however, our models have a significantly higher predictive power since observable properties that are linked in real objects cannot be adjusted independently in the models.

In this paper we restrict the discussion of trends of photometric properties with fundamental stellar and pulsation parameters to a small group of models, in order to present them in some detail. A paper covering photometric results based on a significant fraction of the grid of dynamic model atmospheres presented by Mattsson et al. (\cite{MatWH10}) is currently in preparation. The aim is to make the models available for interpreting of individual stars, as well as for use in stellar evolution and stellar population modelling, by providing a consistent set of wind properties, together with synthetic spectra and photometry for a wide range of parameters. 

The model characteristics and the computational details are outlined in Sect.\,\ref{s:modelling}, while in Sect.\,\ref{s:obsresults} (as well as App.\,\ref{s:compdata}) the comparative data compiled from the literature is presented. In Sect.\,\ref{s:results} we describe how the spectral energy distributions (SEDs) and the photometric variations are influenced by the model parameters, especially by the differing properties of the circumstellar dust envelopes. In Sect.\,\ref{s:comparison} a comparison of various quantities with the corresponding measurements for a sample of galactic Carbon Miras as found in the literature is carried out. Thereafter, we study in Sect.\,\ref{s:discussion} how the models fit into an evolved stellar population (also containing among various red giant stars the targets of our atmospheric modelling, namely C-rich LPVs with mass loss), and discuss the importance of free carbon (C--O) for the dust formation.

\begin{table*}
\begin{center}
\caption{Characteristics of the dynamic model atmospheres for pulsating and mass-losing C-rich AGB stars used for the photometric modelling.}
\begin{tabular}{lll|ccccc}
\hline
\hline
&\multicolumn{2}{l|}{Model:}& T & R & S & C1 & C2  \bigstrut[t]  \\
\hline
(i)&$L_\star$&[$L_{\odot}$]&5200&7000&10\,000&10\,000&10\,000 \bigstrut[t] \\
&$M_\star$&[$M_{\odot}$]&1&1&1&1.5&1.5\\
&$T_\star$&[K]&3000&2800&2600&2400&2400\\
&$[$Fe/H$]$&[dex]&0&0&0&0&0\\
&C/O&\textit{by number}&1.1&1.4&1.4&1.69&2.38 \bigstrut[b] \\
\hline
(ii)&$R_\star$&[$R_{\odot}$]&268&355&493&580&580 \bigstrut[t] \\
&&[\textit{AU}]&1.24&1.65&2.29&2.70&2.70\\
&\multicolumn{2}{l|}{log ($g_\star$ [cm\,s$^{-2}$])}&--\,0.42&--\,0.66&--\,0.94&--\,0.91&--\,0.91 \bigstrut[b] \\
\hline
(iii)&$P$&[d]&295&390&490&640&640 \bigstrut[t] \\
&$\Delta u_{\rm p}$&[km\,s$^{-1}$]&4&5&4&6&6\\
&$f_{\rm L}$& &1&1&2&2&2\\
&$\Delta m_{\rm bol}$&[mag]&0.41&0.53&0.86&1.49&1.49 \bigstrut[b] \\
\hline
(iv)&$\langle\dot M\rangle$&[$M_{\odot}\
$yr$^{-1}$]&--&3.5\,$\times$\,10$^{-6}$&4.3\,$\times$\,10$^{-6}$&9.1\,$\times$\,10$^{-6}$&1.4\,$\times$\,10$^{-5}$ \bigstrut[t] \\
&$\langle u \rangle$&[km\,s$^{-1}$]&--&15&15&17&28.6\\
&$\langle f_c$$\rangle$& &--&0.33&0.28&0.43&0.58\\
&$\langle \rho_{\rm d}/\rho_{\rm g}\rangle$&[10$^{-3}$]&--&0.75&0.63&1.22&3.23 \bigstrut[b] \\
\hline
(v)&$\tau_{\rm dust}$\,(0.55\,$\mu$m)&&--&3.26&4.00&11.06&19.45 \bigstrut[t] \\
&$\tau_{\rm dust}$\,(1.6\,$\mu$m)&&--&1.01&1.24&3.42&6.02\\
&$\tau_{\rm dust}$\,(3.8\,$\mu$m)&&--&0.37&0.45&1.25&2.20 \bigstrut[b] \\
\hline
\end{tabular}
\label{t:dmaparameters}
\end{center}
\textbf{Notes.} Listed are (i) parameters of the hydrostatic initial models, (ii) quantities derivable from these parameters, (iii) attributes of the inner boundary conditions (piston) used to simulate the pulsating stellar interiors as well as the resulting bolometric amplitudes $\Delta m_{\rm bol}$, (iv) properties of the resulting wind, and (v) dust optical depths of the formed circumstellar envelope. 
\newline
The notation follows previous papers (H\"ofner\ et al. \cite{HoGAJ03}; Nowotny et al. \cite{NowHA10}, \cite{NoAHL11}): 
$R_\star$ -- stellar radius of the hydrostatic initial model calculated from the luminosity $L_\star$ and temperature $T_\star$ via the relation $L_\star$\,=\,4$\pi$$R_\star^2$\,$\sigma$$T_\star^4$;
$P$, $\Delta u_{\rm p}$ -- period and velocity amplitude of the piston at the inner boundary; 
$f_L$ -- free parameter to adjust the luminosity amplitude at the inner boundary; 
$\langle\dot M\rangle$, $\langle u \rangle$ -- mean mass-loss rate and outflow velocity at the outer boundary; 
$\langle f_c$$\rangle$ -- mean degree of condensation of the element carbon into amC dust grains at the outer boundary; 
$\langle \rho_{\rm d}/\rho_{\rm g}\rangle$ -- mean dust-to-gas ratio at the outer boundary (cf. Eq.\,(2) of H\"ofner \& Dorfi \cite{HoefD97});
$\tau_{\rm dust}$ -- optical depth of the dust envelope at the specified wavelength (V-, H-, L'-band) based on radial structures at $\phi_{\rm bol}$\,$\approx$\,0.25 (see Sect.\,\ref{s:taudust}).
The radial coordinates in this work are plotted in units of the corresponding stellar radii $R_\star$ of the hydrostatic initial models, calculated from their luminosities $L_\star$ and temperatures $T_\star$ (as given in the table) via the relation $L_\star$\,=\,4$\pi$$R_\star^2$\,$\sigma$$T_\star^4$.
\end{table*}

\section{Model atmospheres and radiative transfer}
\label{s:modelling}

\subsection{Combined atmosphere and wind models}
\label{s:DMAs}

For the study presented here we used the dynamic model atmospheres of H\"ofner et al. (\cite{HoGAJ03}), which cover the two major dynamic aspects relevant to the outer layers of evolved red giants, namely the effects of pulsation and mass loss via a dust-driven wind. To this end, the equations for hydrodynamics, frequency-dependent radiative transfer and dust formation are solved simultaneously. This modelling approach provides a realistic and self-consistent description of the dynamic stellar atmosphere as well as the regions of the outflow, which is triggered by radiation pressure on newly formed amC grains. Such atmospheric models are well suited to simulate mass-losing LPVs with carbon-rich atmospheric chemistry (C/O\,$>$\,1). 

For more information about the numerical and physical details we refer to a number of previous publications dealing with this (e.g. H\"ofner et al. \cite{HoGAJ03}; Gautschy-Loidl et~al. \cite{GaHJH04}; Nowotny et al. \cite{NAHGW05}, \cite{NoLHH05}, \cite{NowHA10}, \cite{NoAHL11}; Mattsson et al. \cite{MatWH10}). Graphical representations of typical models can be found in Nowotny~et~al. (\cite{NowHA10}, \cite{NoAHL11}). These include plots of the temporally varying radial atmospheric structures ($\rho$, $T$, $p$, $u$, $f_c$) at different phases $\phi_{\rm bol}$ during the pulsation cycle and illustrations of the movements of atmospheric layers at different depths. 

In Paper\,II we selected one specific dynamic model atmosphere (named model~S) and studied in detail the influence of circumstellar dust on the photometric appearance as well as the photometric variations throughout a pulsation cycle. This model, also included here (cf. Table\,\ref{t:dmaparameters}), resembles a typical carbon-rich Mira with an intermediate mass-loss rate $\dot M$ of approximately a few 10$^{-6}$\,$M_{\odot}$\,yr$^{-1}$. In addition, we investigate a larger sample of models. The parameters of the respective hydrostatic initial models are listed together with the resultant wind properties in Table\,\ref{t:dmaparameters}.

Although existing objects served as rough reference points for the chosen parameter combinations, the five models should not be regarded as a dedicated fit (for a discussion see Nowotny et al. \cite{NoLHH05}) for any of the targets used for the comparison in Sect.\,\ref{s:comparison}. Instead, the aim was to retrace with the help of characteristic models the sequence of galactic C-type Miras recognisable in observational studies (see below). Therefore, the model parameters were varied to proceed from less to more evolved objects. This means that while the effective temperatures $T_\star$ and the surface gravities $g_\star$ decrease for the model series T\,$\rightarrow$\,R\,$\rightarrow$\,S\,$\rightarrow$\,C1\,$\rightarrow$\,C2, increasing values were chosen for the luminosities $L_\star$, the radii $R_\star$, the C/O ratios, and the pulsation periods $P$. Since one of the significant effects along the observed sequence is the increasingly relevant circumstellar reddening,\footnote{Throughout the paper the term ``reddening'' is used in two different senses: (i) the colour of a star changing due to the dust grains in the \textit{circumstellar} envelope surrounding it, and (ii) the colour becoming redder because of the \textit{interstellar} extinction of the light on its way from the object to us.} we also applied models characterised by increasing mass-loss rates $\dot M$. The latter are dependent on the piston velocity amplitude $\Delta u_{\rm p}$ (see Nowotny et al. \cite{NowHA10}). With appropriately chosen values for these input parameters we can influence the arising outflow to some degree. However, occurrence and intensity of the wind are an outcome of the modelling. In our case, we arrived at models as different as model~T (dust-free, pulsating atmosphere without any outflow) and model~C2 (substantial $\dot M$ exceeding 10$^{-5}$\,$M_{\odot}$\,yr$^{-1}$ and leading to an optically thick circumstellar dusty envelope). They presumably mark extreme values of the mass-loss sequence found observationally, while the previously studied model~S is located roughly in the middle. 

Model~R was adopted from H\"ofner et al. (\cite{HoGAJ03}), while models T and S were taken from Gautschy-Loidl et~al. (\cite{GaHJH04}). The latter model was also used in various other studies (Nowotny et al. \cite{NAHGW05}, \cite{NoLHH05}, \cite{NowHA10}, \cite{NoAHL11}; Paladini et al. \cite{PAHNS09}). Computed specifically for this study were the last two models, C1 and C2.

\subsection{Synthetic spectra and photometry}
\label{s:synthspecphot}

Snapshots of the atmospheric structures at various phases during several periods (see, for example, Fig.\,A.2 in Paper\,II) were then used for the detailed \textit{a posteriori} radiative transfer calculations. To this end we closely followed the numerical approach described in Paper\,II, for the details and an extensive list of references we refer to this previous paper. 

On the basis of the radial structures we applied our opacity generation code {\tt COMA} (Aringer \cite{Aring00}, Paper\,II) to compute atomic and molecular abundances for every depth point of the atmospheric models with equilibrium gas chemistry routines. The principle element abundance pattern of solar composition was modified concerning carbon in two ways. First, the general C abundance was changed according to the C/O ratios as specified for the different models (cf. Table\,\ref{t:dmaparameters}). Second, the depletion of C in the course of amC dust grain formation was also taken into account. Thereafter, opacities for all atmospheric layers and every wavelength point were calculated assuming conditions of LTE. All the sources needed for an appropriate spectral synthesis were included: (i) continuous absorption, (ii) atomic and molecular line contributions, and (iii) opacities due to particles of amC dust under the assumption of the small particle limit. Again we followed the opacity sampling approach of Paper\,II to account for all molecular species necessary for a proper representation of the synthetic spectra (CO, CN, C$_2$, CH, HCN, C$_2$H$_2$, C$_3$). 

Based on the resulting arrays of combined opacities -- with a spectral resolution of $R$\,=\,$\lambda/\Delta\lambda$\,=\,18\,000 --  spherical radiative transfer was solved. The resulting synthetic spectra in the wavelength range of 0.2\,--\,25\,$\mu$m can directly be used to compute filter magnitudes. For the presentation of the spectra in Fig.\,\ref{f:SEDspec} the spectral resolution was decreased to $R$\,=\,360.

Following Paper\,II, we chose for the synthetic photometry the Johnson-Cousins UBVRI system described by Bessell (\cite{Besse90}) as well as JHKLL$^\prime$M from the homogenised Johnson-Glass system described in Bessell \& Brett (\cite{BB88}). The filter transmission curves given by these authors were then convolved with the synthetic spectra. Applying adequate photometric zeropoints for every filter (adopted from Bessell et al. \cite{BesCP98}) led to calibrated synthetic absolute magnitudes. 

Again the Johnson-Cousins-Glass or Bessell photometric system was defined as the standard system for the whole work presented here, and the observational results (Sect.\,\ref{s:obsresults}, App.\,\ref{s:compdata}) were transformed where required, allowing for a direct comparison with the modelling results.

\section{Observational comparative data}
\label{s:obsresults}

As the main observational reference we used the compilation of time-series, near-infrared (NIR) photometry published by Whitelock et al. (\cite{WhFMG06}; below W06). They presented comprehensive data sets for a large sample of galactic carbon-rich AGB stars. Out of the 239 LPVs listed we only adopted the subsample with pronounced variablity, i.e. the well-characterised Miras in their Table\,6. According to Whitelock~et~al. these objects are characterised by clearly periodic lightcurves and peak-to-peak $K$~amplitudes\footnote{In Fig.\,\ref{f:deltaK} the amplitudes of Fourier-fits to the lightcurves (first-order sine curves) are plotted, which can be smaller than the corresponding peak-to-peak amplitudes.} larger than 0.4$^{\rm mag}$, in their nomenclature this corresponds to the variability class 1n. This group of Mira variables is more appropriate for an interpretation by our modelling approach, while the remaining objects (non-Miras, variability class 2n) were not included here. In Paper\,II we made a detailed comparison of the lightcurves of the C-rich Mira RU\,Vir with the corresponding photometric variations of model~S (Table\,\ref{t:dmaparameters}). This is also followed up here to some extent. However, beyond that we made use of the complete sample of observed field C-type Miras. This allows for an extended and more general comparison with models of different parameters. To this end, we adopted the mean values for colour indices from Whitelock et al. rather than the individual measurements of the time-series, NIR photometry also given. The authors provide photometric results already corrected for interstellar reddening as well as useful (estimates for) other quanitities which can be compared to our models. The latter include periods, mass-loss rates $\dot M$, photometric amplitudes $\Delta K$, bolometric magnitudes $m_{\rm bol}$, or distances (which are essential to shift the apparent magnitudes to an absolute scale; e.g. Fig.\,\ref{f:almostHRD}). The photometric results of Whitelock et al. (\cite{WhFMG06}) were not transformed to our standard system (Sect.\,\ref{s:synthspecphot}) as the differences between the SAAO photometric system and the Bessell filters are small according to Bessell \& Brett (\cite{BB88}; see their Table\,I).

We also adopted various other data from the literature for the comparisons below, a detailed description of those including references and the necessary post-processing can be found in App.\,\ref{s:compdata}.

\section{Modelling results -- the influence of increasing mass loss}
\label{s:results}

\subsection{Spectral energy distributions}
\label{s:dusteffects}

In Paper\,II we demonstrated with the help of model~S the difference between hydrostatic objects and dynamic objects (characterised by pulsating stellar interiors and winds) concerning the radial atmospheric structures, the resulting spectra and the photometric appearance. We also outlined the role of different opacity sources for the spectral energy distributions, in particular we showed the effect of the carbon dust particles in the wind. 

\begin{figure}
\resizebox{\hsize}{!}{\includegraphics[clip]{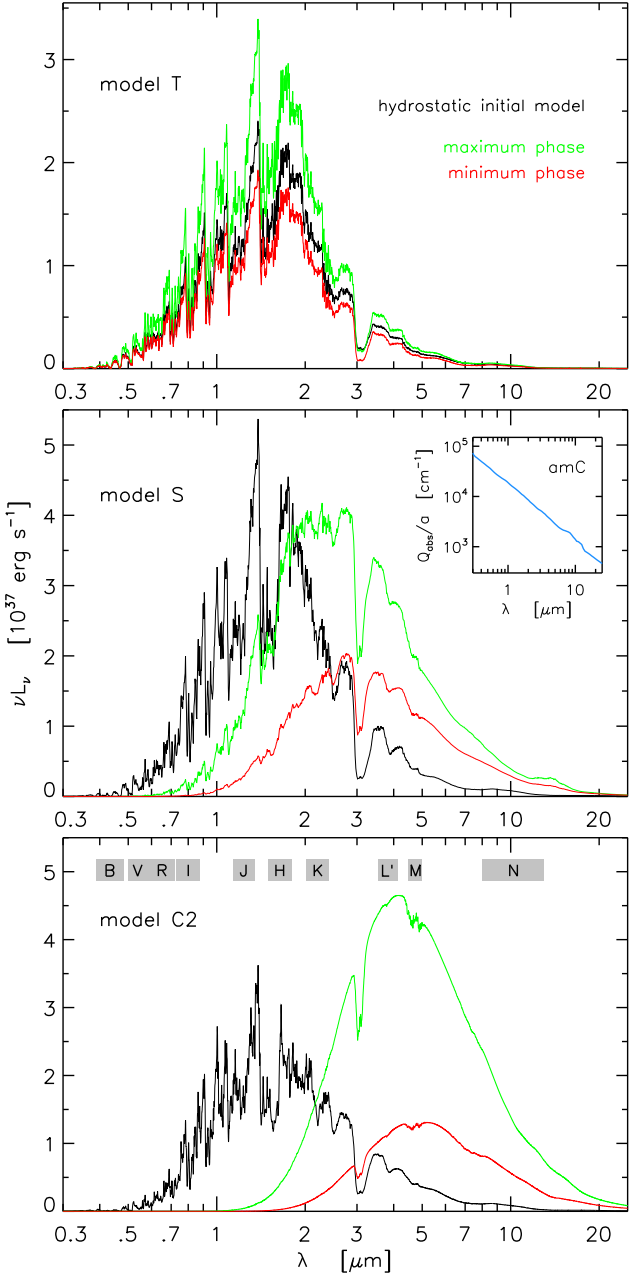}}
\caption{Synthetic low-resolution spectra ($R$\,=\,360) for three dynamic model atmospheres as denoted in the legend. Plotted are spectra based on the corresponding hydrostatic initial model as well as two representative phases of the dynamic calculation ($\phi_{\rm bol} \approx$ {\it 0.0} and {\it 0.5}). The insert in the middle panel shows the absorption data for amorphous carbon dust from Rouleau \& Martin (\cite{RoulM91}) as applied in the modelling. The wavelength ranges of the used broad-band filters ($\approx$\,FWHM of the responses given in Bessell \cite{Besse90} and Bessell \& Brett \cite{BB88}) are marked for orientation purposes in the lower panel.}
\label{f:SEDspec}
\end{figure}

Here we continue the theoretical investigations with the extended set of dynamic model atmospheres of Table\,\ref{t:dmaparameters}. These models have different properties, with one of the most significant being the mass-loss rate. Its influence on the synthetic low-resolution spectra is sketched in Fig.\,\ref{f:SEDspec}, for which we used three models to illustrate the progression from no mass loss to highly dust-enshrouded. For each model we plot the spectra for a maximum and a minimum phase to mark the variations during the pulsation cycle. 

The top panel of Fig.\,\ref{f:SEDspec} shows the results for model~T, which experiences no dust formation and, thus, no stellar wind occurs. The dynamical calculation leads to spectra that are relatively close to the spectrum based on the hydrostatic initial model (which is the starting point for the dynamic calculation; cf. Paper\,II). While all spectra are dominated by the prominent molecular features that are characteristic of late-type giants, the pulsation -- simulated by the varying inner boundary -- results in small variations of the spectrum around the hydrostatic case during a light cycle. In addition to the overall changes in the flux level because of the varying luminosity input, we find changing feature intensities due to the periodically modulated atmospheric structure (cf. Fig.\,3 in Nowotny et al. \cite{NowHA10}).

The corresponding results for model~S are shown in the middle panel of Fig.\,\ref{f:SEDspec}. The stellar wind containing amC grains, which emerges when the conditions for dust condensation are fulfilled, has considerable influence on the spectral appearance of the object. The first recognisable effect is the attenuation of the molecular features in the spectra based on the dynamical phases when compared to the hydrostatic spectrum. The second effect caused by the circumstellar dust is the redistribution of the emitted flux from the visual to the IR. This results in spectra which are clearly different from the hydrostatic one for every phase $\phi_{\rm bol}$ during the whole pulsation cycle.

The bottom panel of Fig.\,\ref{f:SEDspec} displays the dramatic spectral change for even more evolved objects. Model~C2 represents a star with pronounced mass loss leading to an optically thick dust envelope. The molecular contributions in the spectra appear substantially weakened, except for some features (e.g. the characteristic one at $\approx$\,3\,$\mu$m due to C$_2$H$_2$ and HCN), which are still recognisable, although they also have lower intensities. The whole SED is completely shifted to infrared wavelengths and differs significantly from the one for the corresponding initial model (pure photosphere, no wind). In addition, severe changes in flux during the light cycle can be seen.

\begin{figure}
\resizebox{\hsize}{!}{\includegraphics[clip]{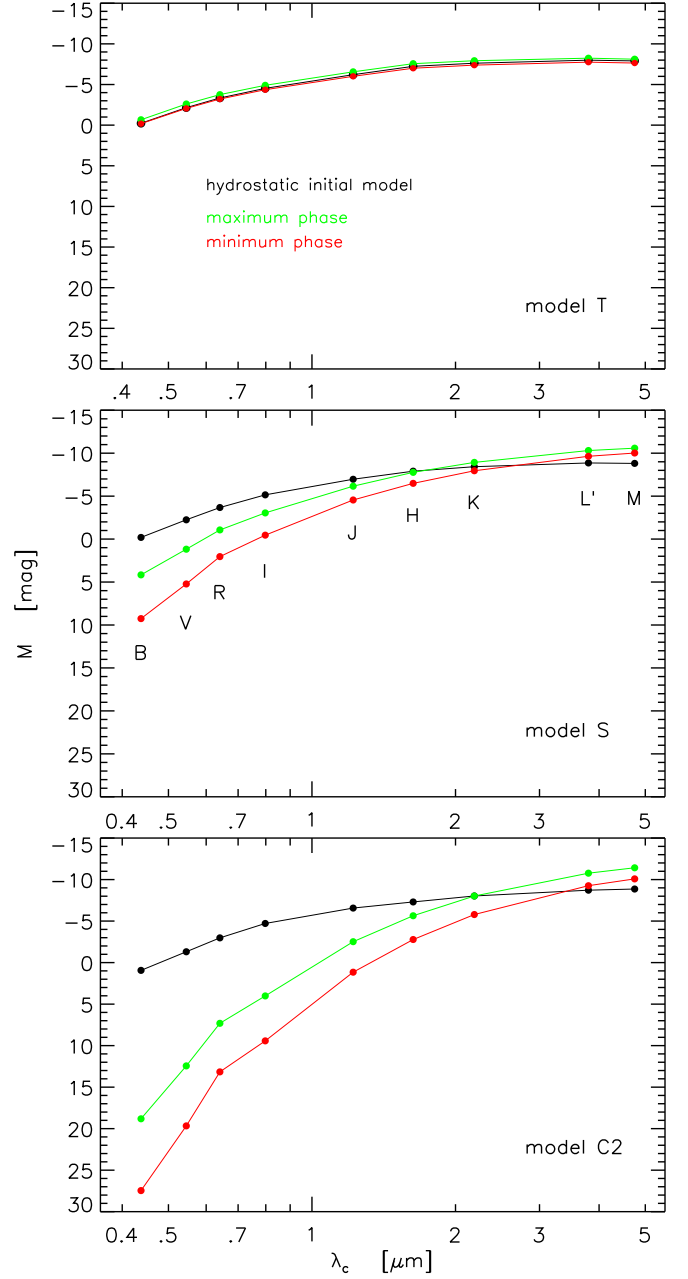}}
\caption{Synthetic SEDs described by calibrated photometric magnitudes in the used broad-band filters resulting from the spectra of Fig.\,\ref{f:SEDspec}.}
\label{f:SEDphot}
\end{figure}

Figure\,\ref{f:SEDspec} illustrates that our modelling approach allows describing carbon-rich LPVs covering a range of quite different mass-loss rates. Starting with objects showing spectra characterised only by photospheric features (atomic, molecular), we proceed to objects for which the spectral appearance is almost completely determined by the circumstellar envelope resulting from the outflow that contains amC dust particles. This effect is also demonstrated in Fig.\,\ref{f:SEDphot} where we show the SEDs of the three selected models of Fig.\,\ref{f:SEDspec} using calibrated filter magnitudes computed on the basis of the corresponding spectra. Only minor deviations ($<$\,1$^{\rm mag}$) from the hydrostatic case can be found for the two dynamic phases of model~T. However, the reddening due to the dust becomes severe for the other models. On the one hand, the objects appear fainter by several magnitudes in the visual due to the absorption caused by the circumstellar dust. The increase in this dimming towards shorter wavelengths reflects the properties of amC dust (cf. insert of Fig.\,\ref{f:SEDspec}) and can be larger than the periodic photometric variations (model~C2). On the other hand, one can also recognise circumstellar dust emission for the filters L$^\prime$M. The change from dimming to additional emission occurs at around 2\,$\mu$m, which is one of the reasons why the $K$ filter is often used in variability studies \mbox{($M_K$-$P$-relation)} of LPVs (e.g. Whitelock \cite{White12}).

\subsection{Optical depths of the dust envelopes}
\label{s:taudust}

\begin{figure}
\resizebox{\hsize}{!}{\includegraphics[clip]{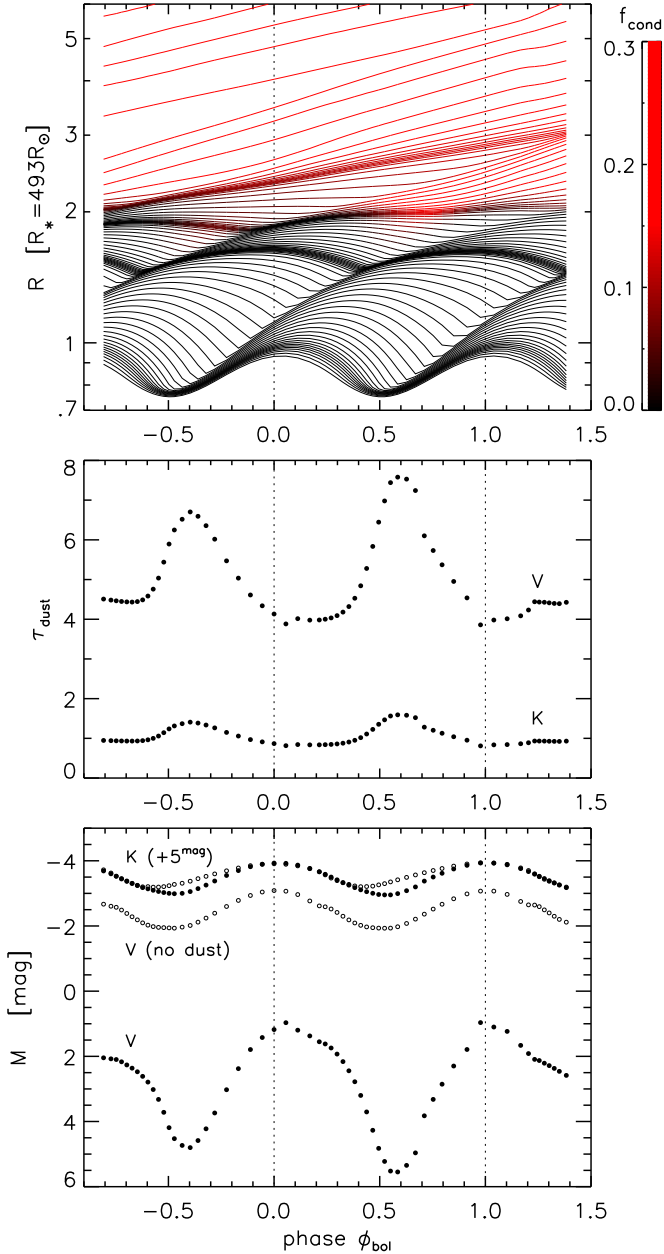}}
\caption{Illustration of the effect of dust formation on the photometric appearance of the star, demonstrated with the help of model~S. \textit{Upper panel:} Movement of atmospheric layers of the model over about two light cycles, colour-coded is the degree of condensation of carbon into dust grains. \textit{Middle panel:} Optical depths of the dust envelope as a function of time, calculated for the central wavelengths of the filters V and K following Eq.\,(\ref{e:optdepth}). \textit{Lower panel:} Resulting synthetic lightcurves for the same filters (filled circles), together with the corresponding photometry for which the dust opacities were not included (open circles). Note that the K-lightcurves were shifted by 5$^{\rm mag}$ to fit into the panel (cf. Fig.\,7 in Paper\,II).}
\label{f:staubillustration}
\end{figure}

To characterise the circumstellar envelopes we computed monochromatic dust optical depths $\tau_{\rm dust}$ at different representative wavelengths by radially integrating inwards from the outer boundary of the model $r_{\rm max}$:
\begin{equation}\label{e:optdepth}
\tau_{\lambda,\rm dust}(r)=\int^{r_{\rm min}}_{r_{\rm max}} \kappa_{\lambda,\rm dust}(r')\,\, \rho(r')\, \mathrm{d}r'
\end{equation}
This corresponds to the central beam with impact parameter $p$\,=\,0 in the framework of spherical radiative transfer (e.g. Fig.\,4 of Yorke \cite{Yorke80}). 

The middle panel of Fig.\,\ref{f:staubillustration} shows how the optical depth of the dusty envelope varies with time for model~S. Superposed on the more or less constant degree of dust absorption due to the steady outflow we also find a temporary increase in $\tau_{\rm dust}$ around phases of light minimum where enhanced dust formation takes place in a relatively narrow region at $\approx$\,2\,$R_\star$ (cf. upper panel of the figure). With the newly emerging dust shell (see also Fig.\,A.2f in Paper\,II), the optical depths peak at $\phi_{\rm bol}$\,$\approx$\,\textit{0.6}. Since the absorption efficiency of amorphous carbon dust has a wavelength dependence of approximately $\propto$\,$\lambda^{-1}$ (see insert of Fig.\,\ref{f:SEDspec}), the effects are more pronounced in the visual than in the NIR. This is also reflected in the lightcurves shown in the lower panel of Fig.\,\ref{f:staubillustration}. There we plot synthetic photometry resulting from the full spectral synthesis (filled circles), as well as the corresponding results of a calculation\footnote{Not consistent with the original dynamical modelling but useful for illustration purposes.} where the dust opacities were not taken into account in the a posteriori radiative transfer (open circles). While the photometric variations of the latter mainly reflect the changing luminosity input introduced by the variable inner boundary of the model (with minor contributions due to varying molecular feature intensities), the former are in addition influenced by the dust effects. For the \mbox{K-filter} only small differences are recognisable at phases where a new dust shell is formed, whereas the \mbox{V-lightcurve} is severly altered by the varying amount of dust absorption. On top of the general dimming by $\approx$\,4$^{\rm mag}$ because of the dusty outflow (visible e.g. at $\phi_{\rm bol}$\,=\,\textit{0.0}) an additional drop in $V$ is clearly noticeable around $\phi_{\rm bol}$\,$\approx$\,\textit{0.6}. In this way, the (time-dependent) absorption by dust grains in the circumstellar envelope is responsible for the generally reddened colours of such objects and also for the variations in colour indices, such as \mbox{($V$\,--\,$K$)}.

Some cycle-to-cycle variations concerning the dust formation and, thus, the V-lightcurve can be recognised in Fig.\,\ref{f:staubillustration}. This effect is relatively moderate for model~S but can be much more pronounced for other parameter combinations (see App.\,\ref{s:double}).

Although we have shown that the dust optical depth is a time-dependent quantity we only give one characteristic number in Table\,\ref{t:dmaparameters}. The values listed there correspond to phases of $\phi_{\rm bol}$\,$\approx$\,\textit{0.25} to avoid the intervals of extreme dust formation discussed above. In this way, the $\tau_{\rm dust}$ should be regarded as lower limits and serve as representative quantites to compare the individual models. We note (i) the monotonic increase of optical depths for the model sequence R\,$\rightarrow$\,S\,$\rightarrow$\,C1\,$\rightarrow$\,C2 (i.e. increasing mass-loss rates), and (ii) the general decrease of $\tau_{\rm dust}$ when going from the visual to the IR (reflecting the absorption properties of amC dust which strictly decrease with wavelength; cf. Fig.\,\ref{f:SEDspec}).

\subsection{Photometric amplitudes in different filters}

\begin{figure}
\resizebox{\hsize}{!}{\includegraphics[clip]{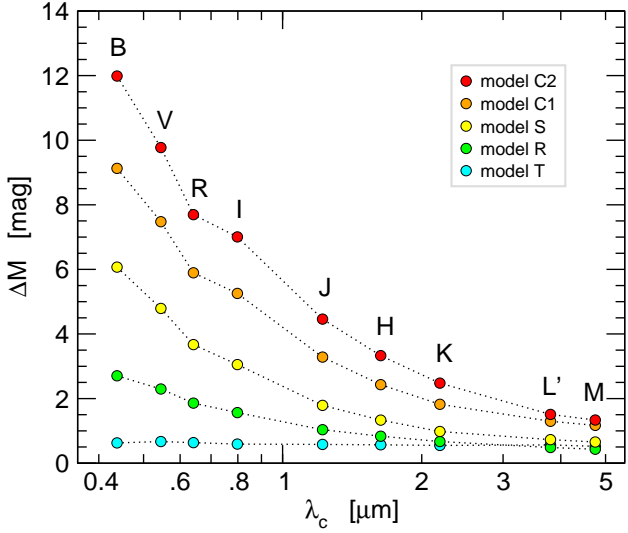}}
\caption{Amplitudes of the photometric variations for each model studied (Table\,\ref{t:dmaparameters}) plotted against the central wavelengths of the various filters.}
\label{f:amplitudes}
\end{figure}

In Paper\,II (see Sect.\,4.1 there) we outlined that for our models dust is \textit{the} decisive factor for the overall shapes of the SEDs and the changing dust absorption also has a strong impact on the resulting photometric amplitude during a lightcycle in the optical. In contrast, the influence of the changed atmospheric structure -- being heavily affected by dynamic processes and, thus, quite different from the hydrostatic case -- on molecular features in the resulting spectra is comparably low. For the set of models in Table\,\ref{t:dmaparameters} we calculated synthetic lightcurves and compared the derived amplitudes in Fig.\,\ref{f:amplitudes}.\footnote{Note that the amplitudes in Fig.\,\ref{f:amplitudes} are computed from the maximum magnitude out of different cycles minus the corresponding minimum value and can therefore be larger than amplitudes estimated from the two extreme phases of the same cycle shown in Fig.\,\ref {f:SEDphot} (cycle-to-cycle variations).} The dust-free model~T shows virtually uniform variations ($\Delta M$\,$\approx$\,0.6$^{\rm mag}$) in all filters. All other models exhibit dusty winds and show the characteristic decrease in amplitudes from the visual towards the IR. This gradient steepens for the sequence R\,$\rightarrow$\,S\,$\rightarrow$\,C1\,$\rightarrow$\,C2. This is partly a consequence of the increase in $\Delta m_{\rm bol}$ (variations of the luminosity input at the inner boundary). However, even more important are the corresponding optical depths of the dusty envelopes. The rising amplitudes in Fig.\,\ref{f:amplitudes} are clearly coupled to the more and more pronounced dust formation along the above-mentioned model sequence. This leads to a general increase in $\tau_{\rm dust}$ (Table\,\ref{t:dmaparameters}), as well as to stronger variations in dust absorption during the light cycle (Sect.\,\ref{s:taudust}).

\section{Testing the models against observations of C-type Miras}
\label{s:comparison}

We calculated similar time series of synthetic spectra and photometry as in Paper\,II for the set of models in Table\,\ref{t:dmaparameters}. However, we restrict the discussion of the modelling results to mean magnitudes and mean colour indices (i.e., averaged over phase $\phi_{\rm bol}$ and over different cycles) for the comparison with observational data of the corresponding Carbon Miras.\footnote{Note that a complete set of lightcurves for all standard broad-band filters of the Johnson-Cousins-Glass system -- as the one for RU\,Vir presented in Fig.\,7 of Paper\,II -- is not available for other C-type Miras.} Also in the current paper we used RU\,Vir as a reference object for model~S and overplot the corresponding data in some of the figures below.

\begin{figure}
\resizebox{\hsize}{!}{\includegraphics[clip]{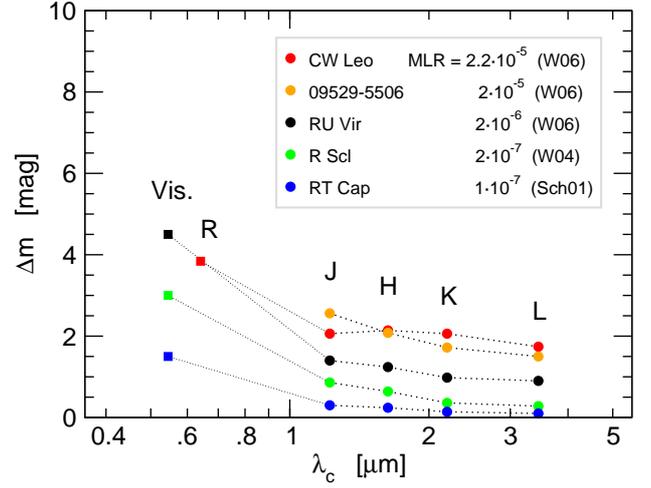}}
\caption{Amplitudes of photometric variations plotted against the central filter wavelengths for a selection of targets monitored by Whitelock~et~al. (\cite{WhFMG06}) in the NIR. The estimates for the mass-loss rates (see legend) were taken from Whitelock et al. (\cite{WhFMG06}), Wong et al. (\cite{WoSLO04}), and Sch{\"o}ier \& Olofsson (\cite{SchoO01}).}
\label{f:amplitudesObs}
\end{figure}

\subsection{Photometric amplitudes in different filters}

To compare these modelling results with observational findings we compiled the data shown in Fig.\,\ref{f:amplitudesObs}. This plot includes a small selection of Carbon Miras from the W06 sample, spanning a representative range in mass-loss rate. While the NIR amplitudes were adopted directly from Whitelock et al. (\cite{WhFMG06}), we had to estimate the values in the optical range from data in the AAVSO\footnote{\tt http://www.aavso.org/data/lcg} database. Such an estimate can be hampered by the irregularities in the lightcurves of observed LPVs. For the characteristic visual amplitudes plotted in Fig.\,\ref{f:amplitudesObs}, we neglected occurring long-term variations (as apparent, e.g. for RU\,Vir in Fig.\,2b of Mattei \cite{Matte97}) that can lead to even higher values ($\Delta m_{\rm vis}$\,$\approx$\,8$^{\rm mag}$ for RU\,Vir). In general, no visual data is available for the objects with large $\dot M$, the amplitude $\Delta R$ of CW\,Leo was taken from the GCVS catalogue (Samus et al. \cite{SDKKP12}). Nevertheless, an overall trend of photometric amplitudes decreasing with wavelength, along with a general increase for more evolved stars (more pronounced pulsation, higher $\dot M$), can be recognised.

\subsection{Colour indices}

\begin{figure}
\resizebox{\hsize}{!}{\includegraphics[clip]{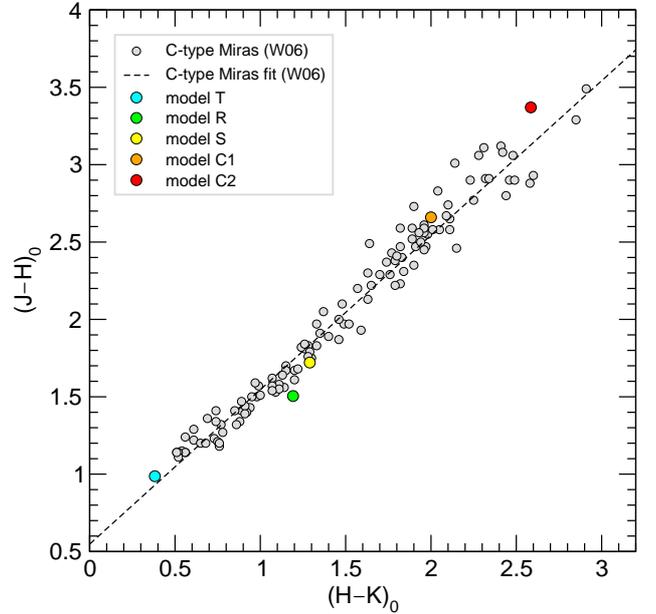}}
\caption{Mean colours corrected for interstellar reddening as listed in Table\,6 of Whitelock et al. (\cite{WhFMG06}) for all C-type Miras. The dashed line represents the fit given in Eq.\,(2) of Whitelock et al. (\cite{WhFMG06}) for the de-reddened Mira sample. Overplotted are the synthetic mean colours of all dynamical models for comparison.}
\label{f:CCDmean}
\end{figure}

Photometric colours are among the most common properties to characterise stellar objects and their overall energy distribution. We start the comparison with the colour-colour diagram based on NIR filter magnitudes shown in Fig.\,\ref{f:CCDmean}. The observed C-rich Miras plotted there cover a large interval in colour, with \mbox{($J$\,--\,$K$)$_{\rm 0}$} ranging from $\approx$\,1.5$^{\rm mag}$ to $\approx$\,6.5$^{\rm mag}$. Whitelock et al. (\cite{WhFMG06}) list a number of even redder stars with ($H$\,--\,$K$)$_{\rm 0}$ very close to $\approx$\,4$^{\rm mag}$; however, such objects lack data in the J-band and cannot be included in Fig.\,\ref{f:CCDmean}. The second reddest object of the well-defined sequence in this plot is the well-known dust-enshrouded carbon star CW\,Leo (or \object{IRC+10216}). To be found at the blue end of the sequence (but not shown here; see Fig.\,15 in Paper\,II) is a number of C-rich LPVs from the W06 sample with less pronounced variability (i.e., most likely SRVs).

Figure\,\ref{f:CCDmean} proves that we are able to trace the whole observational sequence of Miras by applying the different dynamical models. Similar to the hydrostatic COMARCS models, all the corresponding initial models would be located in a quite restricted area in the lower left corner at \mbox{($H$\,--\,$K$)$_{\rm 0}$\,$\approx$\,0.5$^{\rm mag}$} and \mbox{($J$\,--\,$H$)$_{\rm 0}$\,$\approx$\,1$^{\rm mag}$} (not shown here but demonstrated for model~S in Fig.\,16 of Paper\,II), although the stellar parameters of the models (Table\,\ref{t:dmaparameters}) differ considerably. Thus, the dusty winds of different intensities are the main reason for the \mbox{C-type} Miras stretching across the wider colour range in Fig.\,\ref{f:CCDmean} with redder colour indices clearly related to higher mass-loss rates.

\subsection{Mass-loss rates}
\label{s:MLRs}

\begin{figure}
\resizebox{\hsize}{!}{\includegraphics[clip]{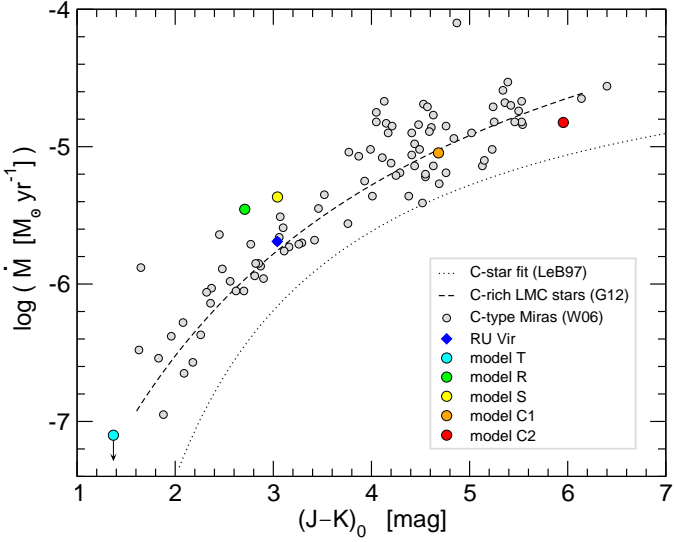}}
\caption{Mass-loss rates $\dot M$ vs. mean dereddened colour indices for the sample of C-type Miras, adopted from Table\,6 of Whitelock et~al. (\cite{WhFMG06}). Also shown are the corresponding results for the dynamical models. For the sake of completeness we mark the location in \mbox{($J$\,--\,$K$)$_{\rm 0}$} of model~T with an arrow although this model does not develop a wind (i.e. $\dot M$\,=\,0). In addition, the empirical relation found by Le~Bertre (\cite{Leber97}) for galactic mass-losing AGB stars of spectral type C is drawn (dotted line). Overplotted is also the more recent fit derived by Gullieuszik et al. (\cite{GGCGL12}) for C-rich giants in the LMC (dashed line).}
\label{f:MLRJK}
\end{figure}

In general, circumstellar dust grains in the outflows of evolved red giants absorb the stellar light in the visual and re-emit it at IR wavelengths, as demonstrated e.g. in Fig.\,\ref{f:SEDspec}. On the basis of this, (dust-) mass-loss rates of such objects can be derived by fitting the redistributed SEDs with the help of dust radiative transfer models (e.g.  Groenewegen et al. \cite{GrSSP09}, Riebel et al. \cite{RiSSM12}). For a review of this topic, a list of modelling approaches, and also the related complications, we refer to van~Loon (\cite{vanLo07}). For a high-quality fit with synthetic SEDs, a sampling of the observed SED with a reasonably large number of representative filters is desirable. This is often not possible, while it is easier to obtain two-colour photometry for the object. Therefore, several studies have tried to establish empirical relations between mass-loss rates $\dot M$ and various colour indices. A comprehensive list of such investigations is given by van~Loon (\cite{vanLo07}), also discussing the limited accuracy of this method for estimating mass-loss rates. 

Nevertheless, comparing models against observational results (e.g. Le~Bertre \cite{Leber97}, Le~Bertre \& Winters \cite{LebeW98}) in a \mbox{$\dot M$-colour} diagram represents an important test of whether the models are adequately describing mass-losing AGB stars. Thus, we carried out such a comparison, as was also done for the Berlin models\footnote{The main difference between their approach and the one used here is that we apply non-grey radiative transfer (including molecular opacities) when computing the dynamic model which has a decisive influence on the radial T-$\rho$-structure. This makes our models suitable to describe spectra and photometry of low-mass-loss objects, in contrast to the Berlin models.} simulating dust-driven winds of LPVs with \mbox{C-rich} (Le~Bertre \& Winters \cite{LebeW98}, Winters et al. \cite{WLJHS00}), as well as \mbox{O-rich} (Jeong et al. \cite{JeWLS03}) atmospheric chemistry. In Fig.\,\ref{f:MLRJK} we plot the mass-loss rates as derived by Whitelock et al. (\cite{WhFMG06}) for their sample of C-type Miras via the mass-loss formula given in Jura (\cite{Jura87}), showing the characteristic rise in NIR colours for increasing $\dot M$. It is interesting that the overplotted empirical relation found for field carbon stars observed by Le~Bertre (\cite{Leber97}; dotted line) lies below the individual data over the whole range. On the other hand, the more recent relation for C-rich LMC stars adopted from Gullieuszik et al. (\cite{GGCGL12}; dashed line) fits the galactic W06 sample better, although these are two independent studies. Anyway, the different dynamic model atmospheres, also shown in Fig.\,\ref{f:MLRJK}, resemble in general the observational sequence of C-type Miras reasonably well.

\begin{figure}
\resizebox{\hsize}{!}{\includegraphics[clip]{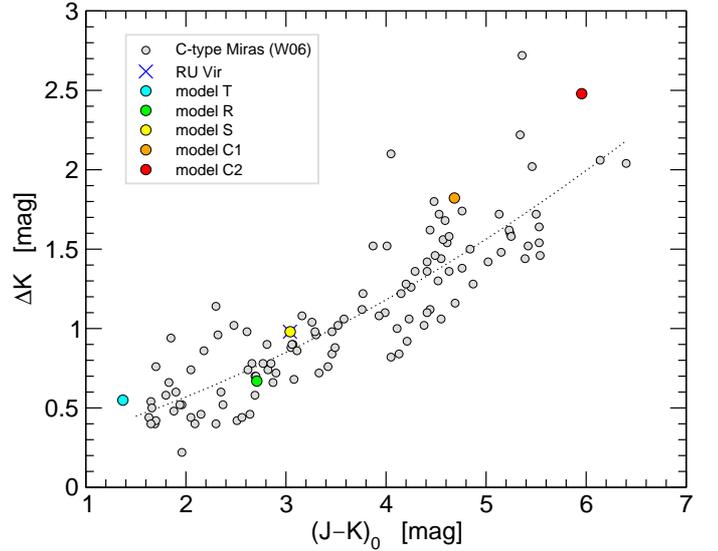}}
\caption{Amplitudes of the $K$-band photometric variations of the C-type Miras listed in Table\,3 of Whitelock et al. (\cite{WhFMG06}) plotted against the corresponding mean colour indices which are corrected for interstellar reddening and given in Table\,6 of Whitelock et al. (\cite{WhFMG06}). The dotted line was fitted to guide the eye. Overplotted are the values resulting from the different atmospheric models. Note, that model~S exactly reproduces the data of RU\,Vir which is, therefore, plotted with a different symbol than in other figures.}
\label{f:deltaK}
\end{figure}

\subsection{Photometric variations}

The effect of circumstellar reddening due to dusty envelopes described above (Fig.\,\ref{f:MLRJK}) can in principle also be reproduced quantitatively by dust radiative transfer codes, such as DUSTY (Ivezi\'{c} \& Elitzur \cite{IvezE97}, Ivezi\'{c} et al. \cite{IveNE99}), the Groenewegen code (Groenewegen \cite{Groen95}, \cite{Groen06}), {\sc 2-Dust} (Ueta \& Meixner \cite{UetaM03}), or  MODUST (Bouwman et al. \cite{BoKAW00}). Based on simplifying assumptions for the circumstellar envelopes (concerning the dust condensation and composition, the velocity law, the radial density structure, the photospheric contribution to the spectra, etc.), these codes compute synthetic SEDs for given sets of input parameters and prescribed radial structures. Although this may provide reasonable relations between colour indices and mass-loss rates, another important aspect can by definition not be reproduced by such an approach, namely the temporal photometric variations. In Fig.\,\ref{f:deltaK}, showing photometric NIR amplitudes against colour indices, all these models calculated with the help of dust radiative transfer codes would line up on the abscissa. In contrast, our models have the advantage of accounting for this effect caused by the pulsating stellar interior. First, they cover the dust envelope \textit{and} the deep photosphere, providing consistent spectra including features due to molecules and dust (Fig.\,\ref{f:SEDspec}). Second, the pulsation triggered in layers below the stellar photosphere is simulated  by the variable inner boundary condition (for details cf. Nowotny et al. \cite{NowHA10}, \cite{NoAHL11}). The latter enables us to calculate synthetic lightcurves and overplot our modelling results in Fig.\,\ref{f:deltaK}. This plot shows the known correlation (Olofsson et al. \cite{OlEGC93}) between photometric variations and circumstellar reddening (i.e., mass-loss rates). The dynamic model atmospheres are able to reproduce objects across the whole range covered by the observational data.

\subsection{Bolometric corrections}
\label{s:BC}

\begin{figure}
\resizebox{\hsize}{!}{\includegraphics[clip]{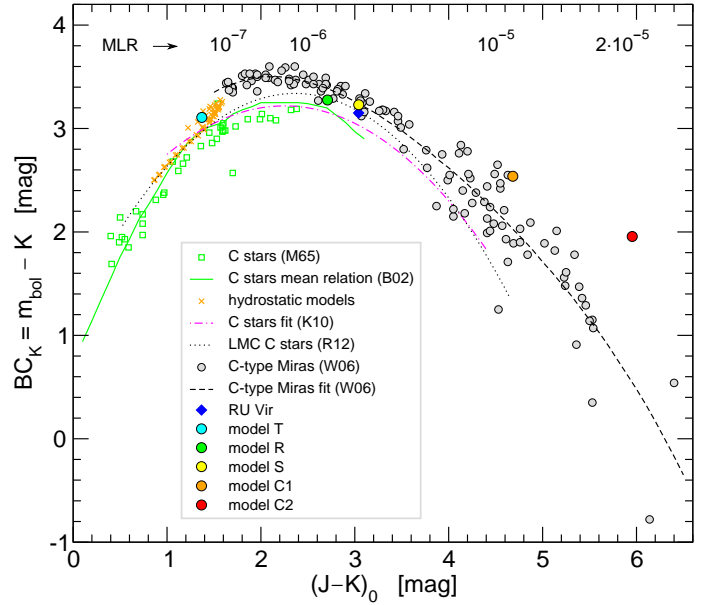}}
\caption{Bolometric corrections $BC_K$ for carbon stars vs. ($J$\,--\,$K$) colours as compiled from different sources: (i) individual measurements for C~stars from Mendoza \& Johnson (\cite{MendJ65}; squares), (ii) the mean relation for \mbox{C-type} giants of Bergeat et al. (\cite{BerKR02}; full line), (iii) synthetic data based on a representative sub-grid of the hydrostatic COMARCS model atmospheres of Paper\,I (crosses), (iv) the fit for nearby field C~stars provided by Kerschbaum et al. (\cite{KerLM10}; dash-dotted line), (v) the fit for C~ star candidates in the LMC derived by Riebel et al. (\cite{RiSSM12}), (vi) individual measurements for field C-type Miras (grey filled circles, blue diamond for RU\,Vir) taken from Whitelock et al. (\cite{WhFMG06}; Table\,6) as well as the corresponding fit given in their Eq.\,(10) and Table\,5 (dashed line). Overplotted are the results for the dynamic model atmospheres of this work. The labels in the upper part mark approximate mass-loss rates $\dot M$~[$M_{\odot}\,$yr$^{-1}$] for the C-type Miras of Whitelock et al. (\cite{WhFMG06}) as related to a given ($J$\,--\,$K$)$_{\rm 0}$ colour in Fig.\,\ref{f:MLRJK}.}
\label{f:BCK}
\end{figure}

In Fig.\,\ref{f:BCK} we tested the atmospheric models in a different way, namely by looking at bolometric corrections (BC) found by observational studies. While there is a body of literature available, we restricted the comparison in the way that (i) only values specifically derived for carbon stars were adopted, and (ii) only data given for the standard JK filters were used. 

The first data set plotted in Fig.\,\ref{f:BCK} are the carbon stars from the sample of Mendoza \& Johnson (\cite{MendJ65}; their Tables\,1 and 2). An empirical fit to these values (not shown in Fig.\,\ref{f:BCK}) was derived by Costa \& Frogel (\cite{CostF96}), given in the CIT/CTIO photometric system used by these authors, though. As outlined in Paper\,I (Sect.\,4.1.4), the hydrostatic COMARCS models -- a representative sample of which is also shown in the figure -- are able to describe the bulk of the Mendoza objects which are not significantly affected by dynamic effects. The mean relation of bolometric corrections given by Bergeat et al. (\cite{BerKR02}; their Table\,A.1) fits the Mendoza objects rather well, even for colours beyond \mbox{($J$\,--\,$K$)$_{\rm 0}$\,$\approx$\,1.6} where the hydrostatic model approach is not adequate anymore.\footnote{The observed targets of Mendoza \& Johnson (\cite{MendJ65}) and Bergeat et al. (\cite{BerKR02}) which have even bluer ($J$\,--\,$K$)$_{\rm 0}$ colours than the COMARCS models of Paper\,I may be C-rich objects which are not classical AGB stars.} 

The relation provided by Kerschbaum et al. (\cite{KerLM10}; their Eq.\,(1)), which rests on blackbody-fits to photometric data, extends to even larger colour indices. As also noted by Kerschbaum et al. (see the discussion in their Sect.\,3.3), there seems to be a general difference to the bolometric corrections derived by  Whitelock et al. (\cite{WhFMG06}) for galactic C-type Miras. Overplotted also in Fig.\,\ref{f:BCK}, the values of the latter authors appear to be most suitable for very red C-rich AGB stars which are seriously influenced by dynamic effects (note that the C-rich LMC objects in Fig.\,12 of Groenewegen et al. \cite{GrSSP09} show a similar distribution). The models presented in this work are constructed to simulate such objects and the corresponding mean values (i.e., averaged over phase $\phi_{\rm bol}$ and over different cycles) can be seen in the plot, too. 

In observational studies the estimation of bolometric magnitudes $m_{\rm bol}$ (and, consequently, $BC$) via integration over photometry in various filters is limited by the available data (which may be the reason behind the above-mentioned differences between different studies, clearly recognisable around \mbox{($J$\,--\,$K$)}$_{\rm 0}$\,$\approx$\,2 in Fig.\,\ref{f:BCK}). On the theoretical side, we have the advantage that the luminosity is an input parameter and bolometric corrections can easily be computed as soon as we have the synthetic photometry. Figure\,\ref{f:BCK} demonstrates that the dynamical models are able to trace the sequence of C-rich objects with increasing mass-loss rates reasonably well. Some differences at the very red end (model~C2) may exist; however, the number of observed objects around ($J$\,--\,$K$)$_{\rm 0}$\,$\approx$\,6 is also limited. In combination with the hydrostatic models the parabolic trend of $BC_K$ vs. ($J$\,--\,$K$)$_{\rm 0}$ for the variety of carbon stars can be followed.

\begin{figure}
\resizebox{\hsize}{!}{\includegraphics[clip]{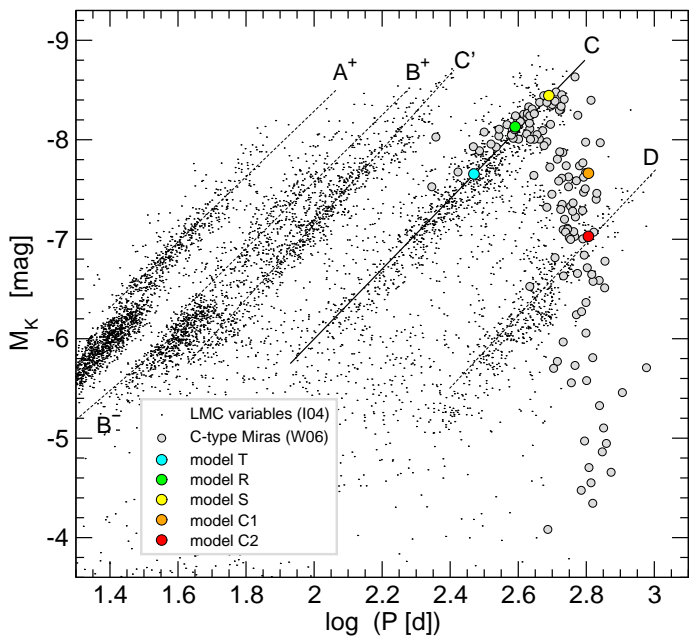}}
\caption{$K$--log($P$) diagram containing the variable stars identified in the LMC by Ita et al. (\cite{ITMNN04a}). Also shown are the $P$-$K$ relations as fitted by the same authors. Both kinds of data were shifted to an absolute scale by assuming a distance modulus \mbox{($m$--$M$)$_{\rm LMC}$} of 18.5$^{\rm mag}$. Overplotted are the C-type Miras adopted from Whitelock et al. (\cite{WhFMG06}; cf. their Tables\,3 and 6) as well as the sequence of dynamic model atmospheres presented in this work.}
\label{f:PLrelation}
\end{figure}

\subsection{Period-luminosity relations}
\label{s:PLR}

A different aspect is examined in Fig.\,\ref{f:PLrelation}, namely the relation between the absolute $K$ magnitudes of evolved red giants and the periods of their lightcurves. It has been known from observational studies of individual objects for quite some time that Mira variables follow a period-luminosity relation (PLR; e.g. Feast et al. \cite{FeGWC89}). Starting with the seminal paper by Wood (\cite{Wood00}) several authors made use of the data sets provided by extended microlensing surveys (e.g. MACHO, OGLE) with the fundamental outcome of LPVs constituting parallel sequences in diagrams as Fig.\,\ref{f:PLrelation}. These investigations were later extended to different stellar systems (e.g. Lorenz et al. \cite{LLNTK11}) and other wavelength ranges (e.g. Glass et al. \cite{GSBSS09}).

For illustration we adopted in Fig.\,\ref{f:PLrelation} the data for LMC variables provided by Ita et al. (\cite{ITMNN04a}). The different sequences which they could identify empiricially are overplotted and labelled in the plot in the same way as given by these authors. The prominent Mira variables, pulsating in the fundamental mode, are located along sequence~C in this plot. Similar \mbox{$K$--log($P$)} relations derived by other authors (e.g. Feast et al. \cite{FeGWC89}, Groenewegen \& Whitelock \cite{GroeW96}, Whitelock et al. \cite{WhiFL08}) are -- within the precision needed for the following comparison with our modelling results -- quite comparable to the fit plotted with a full line. 

It is known that carbon-rich Miras preferentially populate the brighter end of sequence~C (e.g. Wood \cite{Wood00}, Ita et al. \cite{ITMNN04a}). This behaviour can be recognised 
for the C-type Miras of the W06 sample in Fig.\,\ref{f:PLrelation}, at least for the more moderate ones (shorter periods, bluer colours, brighter in K, lower MLRs). Also the models with no or intermediate mass loss (T, R, S) fall right onto sequence~C constituted by Mira variables with moderate outflows. This is not surprising because the stellar parameters (Table\,\ref{t:dmaparameters}) where chosen to follow known PLRs, and the mean $M_K$ is related to $L_\star$ for these objects.

However, there is another eye-catching effect in Fig.\,\ref{f:PLrelation}, which becomes apparent for the more evolved objects. As described by Lattanzio \& Wood (\cite{LattW04}; their Fig.\,2.50) and studied later in detail by Ita \& Matsunaga (\cite{ItaM11}; their Fig.\,2) on the basis of observational data, C-rich Miras with pronounced mass-loss rates show significant deviations from the practically linear relation for fundamental mode pulsators. This effect is wavelength dependent, for the K-band such objects appear clearly fainter (up to 3$^{\rm mag}$) due to the extinction caused by circumstellar dusty envelopes (Sect.\,\ref{s:dusteffects}). The severely reddened targets of the W06 sample in Fig.\,\ref{f:PLrelation} are dimmed by more than 4$^{\rm mag}$ in $K$ and show a distinctive turnoff from sequence~C. Figure\,\ref{f:PLrelation} illustrates that the applied model atmospheres are, in principle, able to reproduce this characteristic behaviour found by observations of \mbox{C-type} Miras. The models resembling more evolved LPVs with longer periods and increasing mass loss (C1, C2) deviate from the linear sequence~C, although they do not become as faint in $M_K$ as the most extreme cases of the observed objects.

\section{Discussion}
\label{s:discussion}

\subsection{The models in the context of a stellar population}
\label{s:almostHRD}

\begin{figure}
\resizebox{\hsize}{!}{\includegraphics[clip]{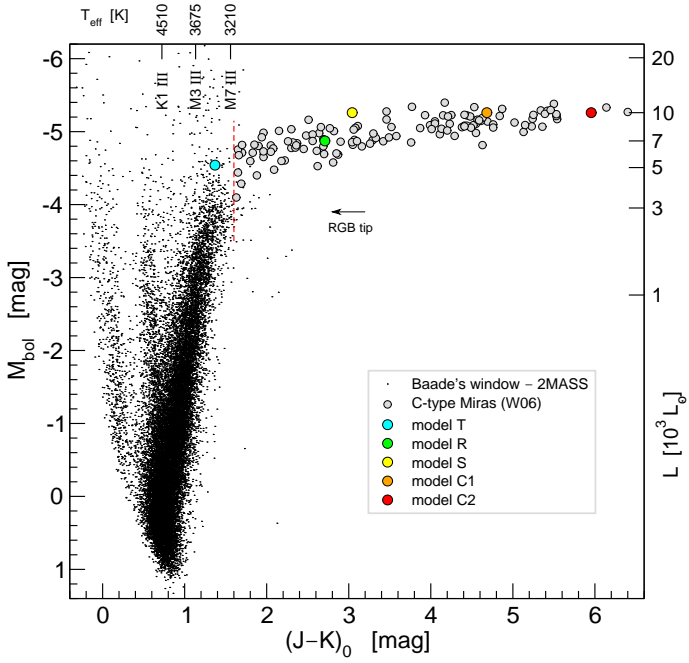}}
\caption{Colour-magnitude diagram containing objects in Baade's window, the bolometric magnitudes were computed from 2MASS photometry by using the $BC_K$ given in Montegriffo et al. (\cite{MoFOF98}). The labels at the top denote effective temperatures and spectral types of late-type giants for a given calibrated colour as listed by Houdashelt et al. (\cite{HoBSW00}, their Table\,4). Overplotted are the corresponding data for the galactic \mbox{C-type} Miras of Whitelock et al. (\cite{WhFMG06}) adopted from their Table\,6, as well as the results for all atmospheric models investigated here. The vertical dashed line at \mbox{($J$\,--\,$K$)\,=\,1.6} marks approximately the border between the hydrostatic and the dynamic regime for C stars (as discussed in the text).}
\label{f:almostHRD}
\end{figure}

To illustrate how the models generally fit into the global picture of a (galactic) stellar population we extracted photometric data from the 2MASS survey for \textrm{Baade's window} (Baade \cite{Baade63}), which is known to be one of the low-extinction windows towards the centre of the Milky Way (e.g. Dutra et al. \cite{DutSB02}). The result of this data mining (cf. App.\,\ref{s:compdata} for details)
is shown in the colour-magnitude diagram (CMD) of Fig.\,\ref{f:almostHRD}. Since the Galactic bulge is dominated by an old, red population, a prominent giant branch can be recognised. It is well populated up to the tip of the red giant branch, which is predicted by theory to be located at $M_{\rm bol}$\,$\approx$\,--3.9$^{\rm mag}$ or $L$\,$\approx$\,2900\,$L_{\odot}$ (e.g. Salaris et al. \cite{SalCW02}, Cassisi \cite{Cassi12}). For orientation purposes the ($J$\,--\,$K$) colours for three representative O-rich objects are indicated at the top of Fig.\,\ref{f:almostHRD}, too, based on the calibrated colour-temperature relation of solar-metallicity models for late-type giants presented by Houdashelt et al. (\cite{HoBSW00}). The sample of Galactic bulge objects\footnote{The old but metal-rich stellar population constituting the Galactic bulge is dominated by stars fainter than the RGB tip, while only a limited number of luminous AGB stars can be found. One would not expect to find carbon stars because of (i) the fact of a small intermediate-age population, in general, and (ii) the relatively high metallicity. A lack of bulge C stars may also be caused by an enhanced oxygen abundance (Feast \cite{Feast07}). Nevertheless, there were studies which claimed to have identified C stars in the bulge (e.g. Azzopardi et al. \cite{AzLRW91}). Appearing quite faint and relatively blue their existence is still puzzling (Whitelock et al. \cite{WhMIF99}, Wahlin et al. \cite{WEGRW07}) and the explanations brought forward range from binary evolution (Whitelock \cite{White93}) to membership of the Sagittarius dwarf galaxy (Ng \cite{Ng98}, \cite{Ng99}). In any case, this group of supposed bulge C stars is consisting of not more than 34 objects. \label{schnucki}} does not contain the kind of stars our models aim to reproduce, namely luminous mass-losing \mbox{C-type} Miras. Therefore, we overplotted in Fig.\,\ref{f:almostHRD} the C-rich field Miras of Whitelock et al. (\cite{WhFMG06}) to complete the overall picture. The latter are among the most luminous stars in this diagram. The plot illustrates two points. First, the luminosities $L_\star$ of the models (input parameter, cf. Table\,\ref{t:dmaparameters}) were chosen close to reality. And second, the models show the same increase in ($J$\,--\,$K$) compared with the sequence of observed \mbox{C-type} Miras. Since this is determined by increased circumstellar reddening, this means that the arising mass-loss rates are also in a realistic range. 

\begin{figure}
\resizebox{\hsize}{!}{\includegraphics[clip]{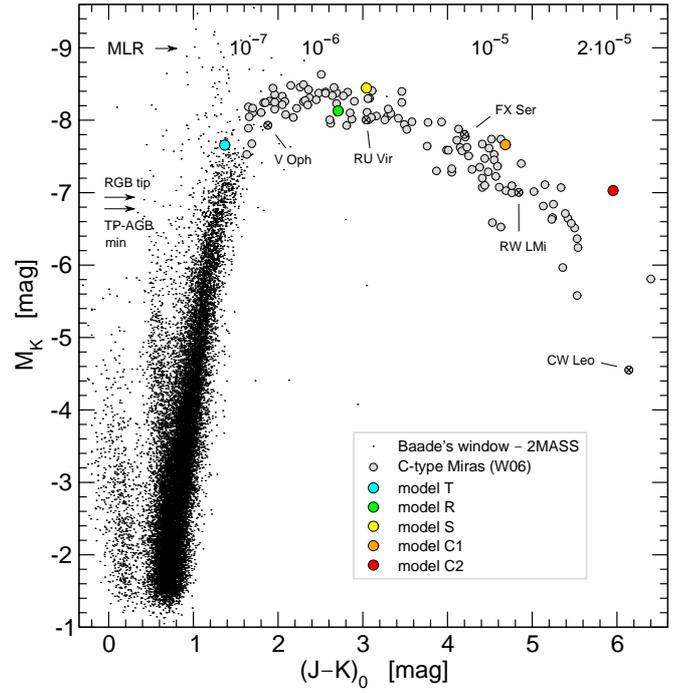}}
\caption{Colour-magnitude diagram of the Galactic bulge objects together with the reddening corrected mean magnitudes and colours of field Carbon Miras from Whitelock et al. (\cite{WhFMG06}, Table\,6) shifted to an absolute scale $M_K$ according to the distances given by the same authors. Overplotted are the mean colours and magnitudes of the series of dynamical models. Marked are, in addition, the magnitude of the RGB tip estimated by using the empirical relations of Valenti et al. (\cite{ValFO04}; Eq.\,(9)) under the assumption of [Fe/H]~$\approx$~0 for the Galactic bulge (Zoccali et al. \cite{ZROGS03}), as well as the magnitude of the beginning of the TP-AGB for a solar-like star as given by Marigo et al. (\cite{MGBGS08}). The labels in the upper part mark approximate mass-loss rates $\dot M$~[$M_{\odot}\,$yr$^{-1}$] for the C-type Miras of Whitelock et al. (\cite{WhFMG06}) as related to a given ($J$\,--\,$K$)$_{\rm 0}$ colour in Fig.\,\ref{f:MLRJK}.}
\label{f:CMDgalacticJKsequence}
\end{figure}

Frequently the members of a stellar population are divided into M- and C-type objects according to a simple colour criterion. Assuming that all stars redder than a certain NIR colour will likely have carbon-rich atmospheric chemistry (Marigo et al. \cite{MarGC03}), C stars are singled out photometrically (e.g. Hughes \& Wood \cite{HughW90}, Wood \cite{Wood00}, Cioni \& Habing \cite{CionH03}, Cioni et al. \cite{CiGMH06}). A typical value applied in studies of stellar systems is \mbox{($J$\,--\,$K$)\,=\,1.4.} However, this has to be regarded as a relatively good indication (e.g. Riebel et al. \cite{RiSSM12}) rather than a strict division in chemistry. On the one hand (comparatively warm) \mbox{C-rich} objects bluer than this supposed borderline were found by observational studies (e.g. the C~stars of Mendoza \& Johnson \cite{MendJ65} in Fig.\,\ref{f:BCK}) as well as in our hydrostatic model grid presented in Paper\,I (all models there have \mbox{($J$\,--\,$K$)\,$<$\,1.6$^{\rm mag}$).} On the other hand there are evolved \mbox{O-rich} objects known which are redder, such as the dust-enshrouded M-type Mira \mbox{IRC-20197} (IW\,Hya), which exhibits according to Le Bertre (\cite{LeBer93}, see their Fig.\,3) a \mbox{($J$\,--\,$K$)} of 3$^{\rm mag}$ and more. Other examples can be recognised among the Galactic bulge Miras of Groenewegen \& Blommaert (\cite{GroeB05}) plotted in Fig.\,\ref{f:CMDgalacticJK}, or in studies of very evolved OH/IR stars (e.g. Marshall et al. \cite{MLMWZ04}). What can be concluded from our modelling, though, is that \mbox{($J$\,--\,$K$)\,$\approx$\,1.6} seems to be the starting point from where the dynamic aspects -- pulsation and  especially the dust effects -- become relevant for C stars (marked with a dashed line in Fig.\,\ref{f:almostHRD}). All of the hydrostatic dust-free COMARCS models of Paper\,I have bluer colours. One of the brightest and nearest optical carbon stars is the extensively studied TX\,Psc which only shows small photometric variability and a weak mass loss (e.g. Gautschy-Loidl et al. \cite{GaHJH04}). Its 2MASS photometry (Cutri et al. \cite{CSDBC03}) transformed using the relations of Carpenter (\cite{Carpe01}) results in \mbox{($J$\,--\,$K$)\,$\approx$\,1.77}, placing TX\,Psc only slightly beyond the assumed border. Clearly redder than \mbox{($J$\,--\,$K$)\,=\,1.6} we find in Fig.\,\ref{f:almostHRD} the Mira variables with pronounced variability and significant stellar winds, both from the observational sample of Whitelock et al. (\cite{WhFMG06}) as well as the corresponding models of our study. Thus, a selection of red stars based on this colour criterion will essentially capture non-hydrostatic, reddened C stars.

Figure\,\ref{f:CMDgalacticJKsequence} shows a similar colour-magnitude diagram for our combined galactic population -- including the bulge objects in Baade's window and the C-rich field Miras adopted from  Whitelock et al. (\cite{WhFMG06}) -- with the dynamic model atmospheres overplotted. In contrast to the previous Fig.\,\ref{f:almostHRD}, where the carbon Miras show luminosities steadily increasing with \mbox{($J$\,--\,$K$)$_{\rm 0}$}, the same objects show here a characteristic dimming in $M_K$ due to the shift of the SED maximum (cf. Sect.\,\ref{s:dusteffects}), reflecting the growing importance of circumstellar dust. Similar CMDs were documented by Zijlstra et al. (\cite{ZMWSL06}) and Lagadec et al. (\cite{LZSMW07}; their Fig.\,2) for the Magellanic Clouds and the C stars observed in these systems with the Spitzer Space Telescope. It can be seen that the models are able to trace the sequence of Carbon Miras from slightly above the RGB tip along the curved branch towards higher ($J$\,--\,$K$)$_{\rm 0}$ colours. This sequence is related to increasing mass-loss rates as denoted in the figure.\footnote{For orientation purposes, a selection of observed targets is also marked in Fig.\,\ref{f:CMDgalacticJKsequence}, with mass-loss rates as derived by Whitelock~et~al. (\cite{WhFMG06}): \object{V Oph} ($\dot M$\,$\approx$\,1.1\,$\times$\,10$^{-7}$\,$M_{\odot}$\,yr$^{-1}$), \object{RU Vir} (2\,$\times$\,10$^{-6}$), \object{FX Ser} (7.6\,$\times$\,10$^{-6}$), \object{RW LMi} (1.1\,$\times$\,10$^{-5}$), and \object{CW Leo} (2.2\,$\times$\,10$^{-5}$).} There may be some deviations in the amount of dimming in $M_K$ when model~C2 is compared to the very reddened observed objects. On the other hand, the observational sequence is also only sparsely populated at ($J$\,--\,$K$)$_{\rm 0}$\,$\approx$\,6$^{\rm mag}$. Further work concerning these discrepancies is required to investigate reasons and remedies.

\subsection{(C--O) versus [Fe/H]}
\label{s:dieCminusOSache}

For the comparison of our modelling results with observational findings we also adopted empirical relations that were originally derived for C-rich objects in the Large Magellanic Clouds (LMC). The first case can be found in Fig.\,\ref{f:MLRJK}, where the models were contrasted with the objects of the W06 sample in the \mbox{$\dot M$-($J$\,--\,$K$)-plane}. Also shown in the figure is the LMC relation derived by Gullieuszik et al. (\cite{GGCGL12}), which fits both kinds of data quite well. The second case concerns the bolometric corrections compiled in Fig.\,\ref{f:BCK}. It appears that the fit derived recently by Riebel et al. (\cite{RiSSM12}) for candidate C-rich AGB stars in the LMC, which is overplotted in Fig.\,\ref{f:BCK}, provides a relatively good overall representation of the bolometric corrections $BC_K$ combined from various data sets (with differences recognisable for the very red objects). 

The agreement between, on the one hand, our models with solar-like composition and the corresponding galactic observational data (esp. the field C-type Miras of Whitelock et al. \cite{WhFMG06}) and, on the other hand, the LMC relations is noteworthy as the metallicities of the Milky Way and the Magellanic Clouds differ (general values of [Fe/H]\,$\approx$\,--0.5 and lower were estimated for the LMC). This suggests that a difference in metallicity is not the decisive factor for mass loss and, thus, the reddening of the objects. In fact, the amount of excess carbon (C--O) -- meaning available C atoms that are not locked in CO molecules -- appears to be more relevant to the dust formation, the properties of the developing wind and the resulting circumstellar reddening. This supports the theoretical results of Mattsson et al. (\cite{MaWHE08}).

\subsection{Extremely reddened C stars}
\label{s:extreme}

Frequently, studies dealing with photometric data obtained for the Magellanic Clouds by the Spitzer SAGE survey (e.g. Vijh et al. \cite{VMBBB09}, Boyer et al. \cite{BSLMM11}) follow the scheme introduced by Blum et al. (\cite{BMOFW06}) to divide the found AGB objects into three distinct classes tagged ``O-rich'', ``C-rich'', and ``Extreme''. Recently, efforts were made to interpret the SEDs of these objects (defined by photometry compiled from various catalogues; cf. Srinivasan et al. \cite{SMLVV09}) with the help of a model grid (Sargent et al. \cite{SarSM11}, Srinivasan et al. \cite{SriSM11}, Riebel et al. \cite{RiSSM12}). This SED fitting led to the conclusion (cf. Fig.\,1 of Riebel et al. \cite{RiSSM12}) that the vast majority of the identified ``extreme AGB stars'' are obscured C-rich objects with high mass-loss rates, while only very few of them might be bright O-rich objects undergoing heavy mass loss (OH/IR stars). In this light, the border between \mbox{``C-rich''} and ``extreme'' AGB stars (set to ($J$\,--\,$K$)\,$\approx$\,2.1$^{\rm mag}$ in Blum et al. \cite{BMOFW06}) appears to be somewhat arbitrary. The latter objects should not be regarded as a distinct class of targets (as suggested by the classification scheme) but rather as C-stars with severe reddening due to circumstellar dust. From the results presented in this work we would assume a natural transition from \mbox{``C-rich''} to ``extreme'' simply caused by increasing mass-loss rates. Such IR-C-stars with colours as red as ($J$\,--\,$K$)\,$\approx$\,6 are also known from previous photometric investigations as, for example, the 2MASS survey of the LMC (Nikolaev \& Weinberg \cite{NikoW00}) or the Galactic bulge (Cole \& Weinberg \cite{ColeW02}).

\section{Summary and conclusions}
\label{s:summary}

The aim of this work was to reproduce the observable variety of mass-losing galactic Miras of carbon-rich atmospheric chemistry with the help of sophisticated model atmospheres. In particular, we wanted to see if our modelling approach is able to simulate the photometric properties of the comprehensive sample of well-characterised \mbox{C-type} field Miras observed by Whitelock et al. (\cite{WhFMG06}). These objects form -- more or less pronounced -- sequences in various kinds of diagrams. Several properties characterising the stars (e.g. luminosities, colours, pulsation periods, photometric amplitudes) show on average an increasing trend along the sequences. This could suggest an evolution of C-Miras along this series of observed targets. However, one should keep in mind that the W06 sample of field Miras might contain a mixture of different masses and ages, making the interpretation as an direct evolutionary sequence arguable. Nevertheless, a general trend from less to more evolved objects is recognisable, with the increasing mass loss being one of the most significant effects causing circumstellar reddening.

For our photometric modelling we used time-dependent dynamic model atmospheres that simulate the dynamic photospheres, as well as the pulsation-enhanced dust-driven winds of C-rich LPVs (H\"ofner\ et al. \cite{HoGAJ03}, Mattsson et al. \cite{MatWH10}). It should be noted that this self-consistent approach differs from frequently used dust radiative transfer codes to model dust envelopes of mass-losing AGB stars. One major difference is the fact that the development of a stellar wind is the result of the chosen set of fundamental parameters, i.e., the mass-loss rate $\dot M$ is not an input parameter but actually an outcome of the modelling. Another advantage is that we are also able to compute variations in spectroscopic and photometric properties, which is crucial for intrinsically variable objects such as Miras. 

A set of five models with quite different parameters was used to represent the diversity of Carbon Miras found in observational studies. Based on the radial atmospheric and wind structures we computed (consistent with the preceding modelling) synthetic spectra and photometry for standard broad-band filters. In a first step we presented low-resolution spectra for different models, illustrating that our modelling method is able to describe a diversity of objects ranging from only moderately pulsating LPVs without any dusty outflow to evolved Miras with pronounced variability and optically thick circumstellar dust envelopes. In this way, our modelling approach bridges the gap between (i) the Australia-Heidelberg-type models for dust-free pulsating atmospheres without wind, and (ii) the Berlin-type models for dust-driven winds of highly dust-enshrouded objects (see Paper\,II and Sect.\,3 of Nowotny et al. \cite{NAHGW05} for a more detailed description and references). Our models can reproduce spectra of C-rich AGB stars, for which the respective contributions by the photosphere and the dusty envelope are of varying relevance. This also includes objects with significant mass loss but no optically thick winds.

The applied models compare well to the observational data and can (quantitatively and simultaneously) reproduce the observed values and ranges for absolute magnitudes, colour indices, photometric variations, mass-loss rates, and bolometric corrections. We were able to retrace the sequence of galactic \mbox{C-type} Miras from the ``blue end'' (stars slightly beyond the tip of the RGB, modest variations, no outflow) to the ``red end'' (evolved red giants, large photometric amplitudes, pronounced circumstellar reddening due to heavy mass loss) in diverse diagrams: CMDs, two-colour diagrams, $\dot M$ vs. \mbox{($J$\,--\,$K$)}, $\Delta K$ vs. \mbox{($J$\,--\,$K$)}, or $K$--log($P$) diagrams. From this we can conclude that (i) the input parameters of the models were chosen realistically (e.g. $L_\star$\,+\,$\Delta u_{\rm p}$\,+\,$f_{\rm L}$ $\rightarrow$ $\Delta K$ reproduced), and (ii) the stellar winds emerging in the modelling scenario have properties resembling well the corresponding observed objects (e.g. $\dot M$\,+\,$u_{\rm exp}$\,+\,$f_c$ $\rightarrow$ red \mbox{($J$\,--\,$K$)} colours reproduced), suggesting an adequate modelling procedure. The combination of both enables us to reproduce the diversity of observed populations of \mbox{C-rich} Miras.

\begin{acknowledgements}
This research was funded by the Austrian Science Fund (FWF): P21988-N16, P23006, P23586.
This work was supported by the Swedish Research Council. 
BA acknowledges funding by the contracts ASI-INAF I/016/07/0 and ASI-INAF I/009/10/0. 
Sincere thanks go to T. Lebzelter and J. Hron for careful reading and fruitful discussions.
We acknowledge with thanks the variable star observations from the AAVSO International Database contributed by observers worldwide and used in this research.
This research has made use of 
(i) NASA's Astrophysics Data System, 
(ii) the VizieR catalogue access tool, CDS, Strasbourg, France,
(iii) the SIMBAD database, operated at the CDS, Strasbourg, France, and
(iv) the NASA/IPAC Infrared Science Archive, which is operated by the Jet Propulsion Laboratory, California Institute of Technology, under contract with the National Aeronautics and Space Administration.
This publication makes use of data products from the Two Micron All Sky Survey, which is a joint project of the University of Massachusetts and the Infrared Processing and Analysis Center/California Institute of Technology, funded by the National Aeronautics and Space Administration and the National Science Foundation.
We thank the anonymous referee for the comments that helped improve the paper.
\end{acknowledgements}

\Online

\begin{appendix}

\section{Various comparative data}
\label{s:compdata}

In Sect.\,\ref{s:almostHRD} (and App.\,\ref{s:loopsCMDs}) we use photometry of the 2MASS All-Sky Point Source Catalog (Cutri et al. \cite{CSDBC03}) for a section of the Galactic bulge, which is available from the NASA/IPAC Infrared Science Archive (IRSA).\footnote{\tt http://irsa.ipac.caltech.edu/} Photometric data were downloaded for objects of a field located in \textit{Baade's window}, we extracted photometry for an area of $\Delta\alpha\times\Delta\delta$~$\approx$~0.5$^\circ\times$~1$^\circ$. Only stars fulfilling $\sigma_{JHK}$~$<$~0.05$^{\rm mag}$ were chosen, the correction for interstellar reddening was carried out according to the extinction values given by Dutra et al. (\cite{DutSB02}). The 2MASS photometry had to be transformed to our Bessell system using the equations given in Carpenter (\cite{Carpe01}; see their App.\,A). To convert the $K$-magnitudes of 2MASS to bolometric magnitudes (Fig.\,\ref{f:almostHRD}) we  made use of the bolometric corrections $BC_K$ as a function of ($J$\,--\,$K$) derived by Montegriffo et al. (\cite{MoFOF98}). To this end we fitted the values tabulated in Montegriffo's Table\,3, which are labelled `metal-rich stars'. The resulting regression\footnote{$BC_K$ = 0.148 + 4.545 $\cdot$ ($J$\,--\,$K$) -- 2.786 $\cdot$ ($J$\,--\,$K$)$^2$ + 0.786 $\cdot$ ($J$\,--\,$K$)$^3$} was then applied to the 2MASS data.

In addition, we used a sample of O-rich Miras in the Galactic bulge identified by Groenewegen \& Blommaert (\cite{GroeB05}) while investigating the lightcurves produced by the OGLE-II survey. The authors cross-correlated their objects with the 2MASS catalogue leading to a subsample of 1619 LPVs which are plotted in Fig.\,\ref{f:CMDgalacticJK}. Dereddening was done following the values for interstellar extinction for each OGLE-II field as listed in Matsunaga et al. (\cite{MatFN05}). For the conversion between 2MASS and Bessell filters we applied again the relations of Carpenter (\cite{Carpe01}).

The reddening-corrected photometric data (apparent magnitudes) for all bulge objects were shifted to absolute magnitudes by assuming a distance modulus of ($m$\,--\,$M$)~=~14.7$^{\rm mag}$ for the centre of our galaxy (e.g. Vanhollebeke et al. \cite{VanGG09}).

For the comparison in Fig.\,\ref{f:PLrelation} we made use of the data for variable stars in the Large Magellanic Cloud (LMC) presented in Ita et al. (\cite{ITMNN04a}). The authors analysed OGLE lightcurves and provide estimates for the periods for a large number of objects, as well as the corresponding JHK photometry. We dereddened the data with the value for $A_{\rm K}$ given by Feast et al. (\cite{FeGWC89}) and transformed the near-IR photometry from the LCO to the Bessell system with the help of Carpenter (\cite{Carpe01}). We also adopted the $P$-$K$ relations fitted by Ita et al. (\cite{ITMNN04a}; cf. their Table\,3) and overplotted them in Fig.\,\ref{f:PLrelation}. As a distance modulus for the LMC we assumed ($m$\,--\,$M$)~=~18.5$^{\rm mag}$ (e.g. Gibson \cite{Gibso00}, Freedman et al. \cite{FMGFK01}, Alves \cite{Alves04}, Laney et al. \cite{LanJP12}). 

The relation between mass-loss rates $\dot M$ and \mbox{($J$\,--\,$K$)} colours empirically found by Le~Bertre (\cite{Leber97}) and plotted in Fig.\,\ref{f:MLRJK} is given in the ESO photometric system and was transformed to the Johnson-Glass system according to the equations given in Bessell \& Brett (\cite{BB88}). The other relation given in Gullieuszik et al. (\cite{GGCGL12}; their Eq.(2)) was transformed from the 2MASS system by applying the conversion of Carpenter (\cite{Carpe01}).

The magnitudes of the solar-like hydrostatic COMARCS models -- a sub-grid is used in Fig.\,\ref{f:BCK} -- are of the same kind as the corresponding ones of the dynamic models (Paper\,I). 

Finally, we compiled various data sets from different sources for the comparison of bolometric corrections (BCs) shown in Fig.\,\ref{f:BCK}. The observational results of Mendoza \& Johnson (\cite{MendJ65}) are already given in the Johnson system, no conversion is needed therefore. The fit to the values of an extended sample of C-type giants provided by Bergeat et al. (\cite{BerKR02}; their Table\,A.1) is based on observed data collected by the authors from the literature. According to Knapik et al. (\cite{KnapB97}) these measurements were obtained in the "Arizona system or close to it" (i.e., Johnson) and can thus be compared directly (Bessell \& Brett \cite{BB88}, Table\,I) to the models. The fit to the data of a sample of nearby field C stars provided by Kerschbaum et al. (\cite{KerLM10}) in their Eq.\,(1) is based on photometry in the new ESO system, which is very close to the Johnson system according to the authors (see also Bouchet et al. \cite{BouSM91}). The fit for C-rich AGB candidates in the LMC given by Riebel et al. (\cite{RiSSM12}) in their Table\,7 was transformed from the 2MASS system to our standard system with the help of Carpenter (\cite{Carpe01}).

\end{appendix}

\begin{appendix}

\section{Loops in colour-magnitude diagrams}
\label{s:loopsCMDs}

In the previous Sects.~\ref{s:comparison} and \ref{s:discussion} we analysed the applied models on the basis of mean photometry. Here we want to outline how typical C-type Miras vary within CMDs constructed from NIR photometry.

\begin{figure}
\resizebox{\hsize}{!}{\includegraphics[clip]{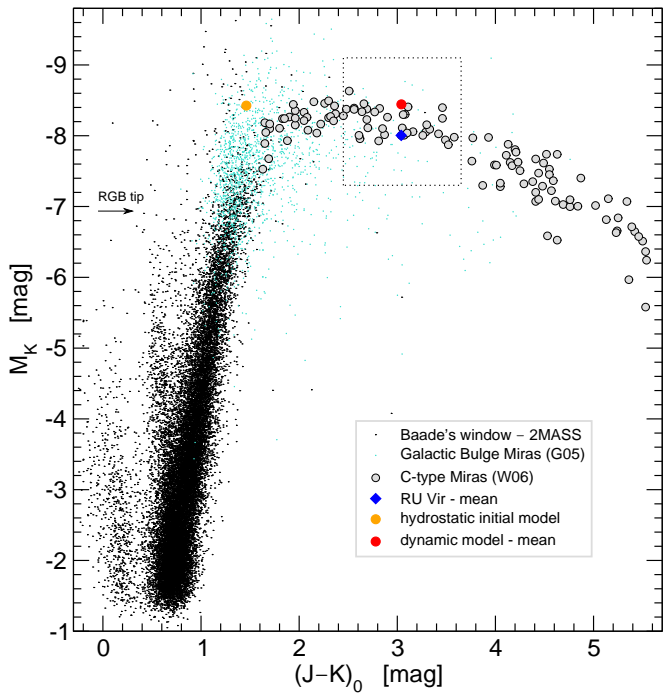}}
\caption{Colour-magnitude diagram containing observational data compiled from different sources: (i) 2MASS photometry of stars in Baade's window (black dots), (ii) 2MASS photometry of Mira variables in the Galactic bulge (turquoise dots) identified by Groenewegen \& Blommaert (\cite{GroeB05}), and (iii) reddening corrected mean magnitudes and colours of field C-Miras (grey filled circles, blue diamond for the individual RU\,Vir data) from Whitelock et al. (\cite{WhFMG06}, Table\,6) shifted to an absolute scale $M_K$ according to the distances given by the same authors. Overplotted are the modelling results for model~S, i.e. the location of the hydrostatic initial model as well as the average of several phases of the different cycles in Fig.\,\ref{f:CMDDMARUVir}. The colour-code is the same as in Fig.\,16 of Paper\,II. The box (dotted lines) marks the range covered in Fig.\,\ref{f:CMDDMARUVir}.}
\label{f:CMDgalacticJK}
\end{figure}

In Fig.\,\ref{f:CMDgalacticJK} we show the same combined population of galactic objects as in Fig.\,\ref{f:almostHRD}, but with absolute $K$ magnitudes on the ordinate instead of bolometric ones. This enables us to overplot also the Miras (most likely \mbox{O-rich}$^{\ref{schnucki}}$) found by Groenewegen \& Blommaert (\cite{GroeB05}) in the 49 OGLE fields across the Galactic bulge. For a description of the necessary post-processing of the 2MASS photometry we refer to App.\,\ref{s:compdata}. The fact that the 2MASS survey provides `only' single-epoch data (instead of a time series that could be averaged) introduces some scatter in for these variable stars. Still it is clear that the latter objects are all located on the extension of the pronounced red giant branch towards even brighter $M_K$ in Fig.\,\ref{f:CMDgalacticJK}. The C-type Miras of the W06 sample form again the noticeable sequence from the bright end of the red giant branch towards redder colours with ($J$\,--\,$K$)$_{\rm 0}$ up to $\approx$\,6$^{\rm mag}$. Also plotted in Fig.\,\ref{f:CMDgalacticJK} is the synthetic photometry for model~S, illustrating the difference between the dust-free hydrostatic initial model and the developed dynamic model in a CMD.

\begin{figure}
\resizebox{\hsize}{!}{\includegraphics[clip]{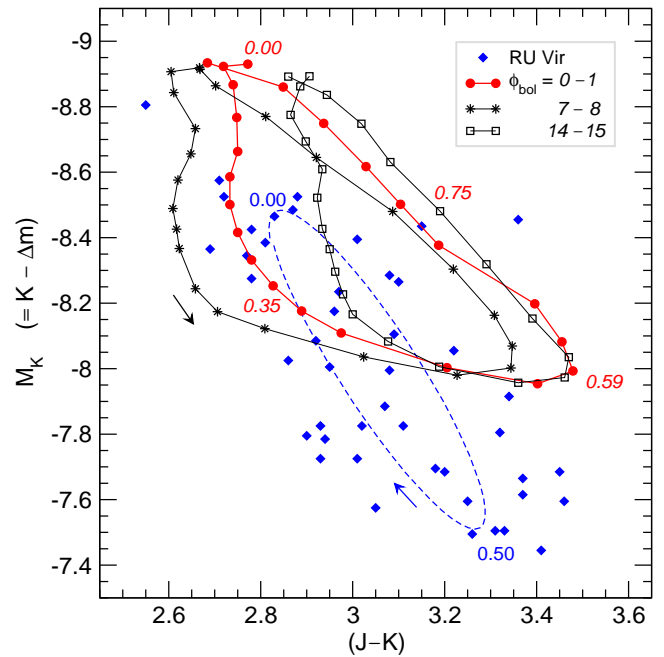}}
\caption{Colour-magnitude diagram containing individual observations of RU\,Vir adopted from Table\,2 of Whitelock et al. (\cite{WhFMG06}) and shifted to an absolute scale $M_K$ by applying the distance modulus derived by the same authors (cf. their Table\,6). The dashed line marks the average variation of RU\,Vir during the light cycle as derived from the Fourier fits in Fig.\,8 of Paper\,II. Overplotted are the corresponding synthetic photometric data of model~S. Three different pulsation cycles are plotted with the same symbols, colour-coding, and labels as in Fig.\,13 of Paper\,II. The arrows mark the directions of how the objects pass through the loops during a light cycle, selected phases are labelled (convention for $\phi_{\rm bol}$ / $\phi_{\rm v}$ as in Paper\,II).}
\label{f:CMDDMARUVir}
\end{figure}

Figure\,\ref{f:CMDDMARUVir} (showing a close-up of Fig.\,\ref{f:CMDgalacticJK}) illustrates the temporal variations, with model~S and the C-type Mira RU\,Vir as examples. Both occupy the same range of the CMD and show variations on the order of 1$^{\rm mag}$ in $K$ as well as \mbox{($J$\,--\,$K$)$_{\rm 0}$}. Apart from the recognisable cycle-to-cycle variations the model shows for every pulsation period similar broad loops throughout the light cycle. It appears bright and blue at maximum light (phase $\phi_{\rm bol}$\,$\approx$\,0.0), while it is faint and red around minimum light ($\phi_{\rm bol}$\,$\approx$\,0.5). As described in Paper\,II, we cannot plot such loops for RU\,Vir directly as the observed data points were obtained with an insufficient sampling rate over several periods. Instead, we merged the measurements from different periods into one combined light cycle. Based on the sine fits (one Fourier component) to the resulting lightcurves (see Fig.\,8 in Paper\,II) we can derive \textit{simulated} variations of RU\,Vir, which are shown with a dashed line in Fig.\,\ref{f:CMDDMARUVir}. The range of this simulated loop is narrower than the range covered by the individual data points. As noted in Paper\,II (cf. the comparison of original \mbox{($J$\,--\,$K$)} data vs. the corresponding simulated colour curve in Fig.\,12 there), this can be explained by the occuring cycle-to-cycle variations in the lightcurves. Nevertheless, the general behaviour of looping through the CMD, with extreme phases in the same order ($\phi_{\rm bol}$\,$\approx$\,0.0 at the upper left end, $\phi_{\rm bol}$\,$\approx$\,0.5 at the lower right end), is similar to the model. Again we find that the model loops counter-clockwise in the CMD of Fig.\,\ref{f:CMDDMARUVir}, while RU\,Vir follows a clockwise movement. We demonstrated in the appendix of Paper\,II how a (relatively small) phase shift between the lightcurves obtained in the different filters can lead to a change in the sense of rotation. Interestingly, this behaviour has only scarcely been investigated directly by observational studies. Payne-Gaposchkin \& Whitney (\cite{PaynW76}; their Figs.\,5abc) studied a large sample of LPVs and found them -- averaged over certain groups -- to vary in CMDs very similarly to our results. Unfortunately, they show only loops for stars of spectral types M and S (clockwise), while the corresponding data for C-rich objects was according to the authors afflicted with large scatter. Eggen (\cite{Eggen75}; see his Figs.\,18, 24, and 30) monitored large-amplitude variables (mostly M-type) and plotted their variations in the \mbox{$M_{\rm bol}$-($R$\,--\,$I$)-plane}. Although the sampling of these time series is limited, the stars seem to perform loops with the sense of rotation being mostly counter-clockwise. A slightly different approach was pursued by Wing (\cite{Wing67}), who obtained a spectro-photometric time series over $\approx$\,2.5 years for about 25 Miras from which he could derive oxide band strengths and temperature estimates. According to Wing (priv.\,comm.) the objects showed quite a variety of behaviours. The results for only five of these objects were then published in Spinrad \& Wing (\cite{SpinW69}; their Fig.\,9), showing clockwise as well as counter-clockwise variations.\\

Note that in Fig.\,3 of Nowotny et al. (\cite{NoAHL11b}) we plotted a colour-magnitude diagram $M_K$ vs. ($V$\,--\,$K$)$_{\rm 0}$, showing the combined contents of Figs.\,\ref{f:CMDgalacticJK} and \ref{f:CMDDMARUVir} from here. Although this differing filter combination provides a valuable colour index because of the broad basis in wavelength, it is generally harder to obtain simultaneous measurements in these two filters. For most of the objects in the W06 sample there is no information concerning their brightnesses or even light variations in the \mbox{V-band} and they are not included in this plot. For a quite limited sub-sample of the C-type Miras we could estimate a mean visual magnitude from the photometric time series available in the AAVSO database, providing a hint on a mean \mbox{($V$\,--\,$K$)$_{\rm 0}$.} This is only possible for stars that are bright in the visual and not strongly influenced by circumstellar reddening. Thus, only objects with $\dot M$\,$\leq$\,2\,$\times$\,10$^{-6}$\,$M_{\odot}$yr$^{-1}$ could be studied. The C-rich Mira RU\,Vir is the only target for which simulated variations in the \mbox{$M_K$-($V$\,--\,$K$)$_{\rm 0}$-plane} could be derived from the Fourier fits of the lightcurves (see Fig.\,8 in Paper\,II). Both RU\,Vir and model~S show a relatively similar behaviour as found here (loops), with the averaged colour of the latter being significantly redder, though.

\end{appendix}

\begin{appendix}

\section{Extreme dust formation}
\label{s:double}

\begin{figure}
\resizebox{\hsize}{!}{\includegraphics[clip]{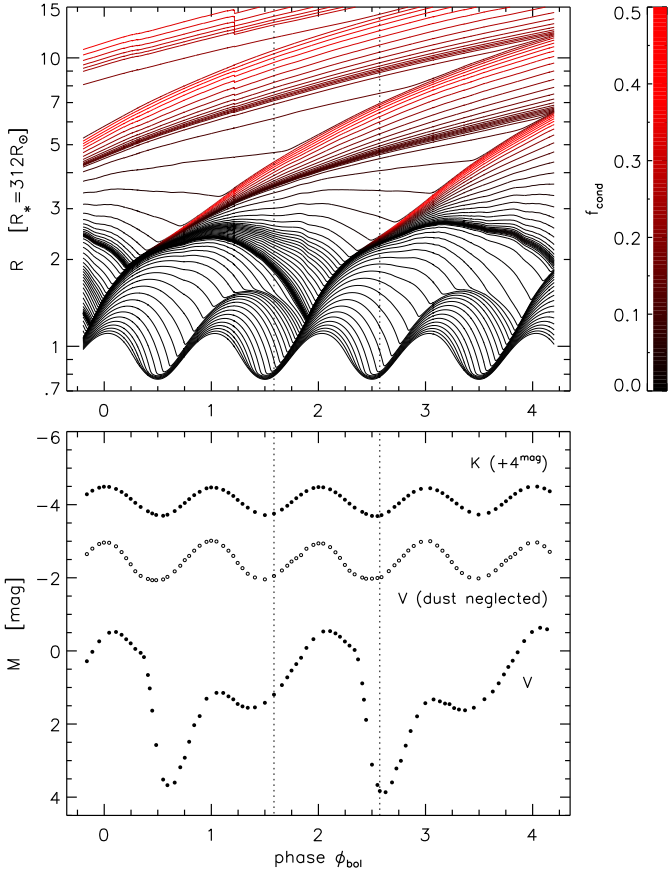}}
\caption{\textit{Upper panel:} Movement of the outer layers of a specific dynamic model atmosphere with dust formation occuring every second pulsation period (App.\,\ref{s:double}). Colour-coded is the degree of condensation of carbon atoms into amorphous carbon dust particles. \textit{Lower panel:} Synthetic lightcurves for the broad-band filters V and K (filled circles), together with the corresponding photometry for which the dust opacities were neglected (open circles). Note that the \mbox{K-lightcurves} were shifted by 3$^{\rm mag}$ to better fit into one panel.}
\label{f:doubleshellphot}
\end{figure}

\begin{figure}
\resizebox{\hsize}{!}{\includegraphics[clip]{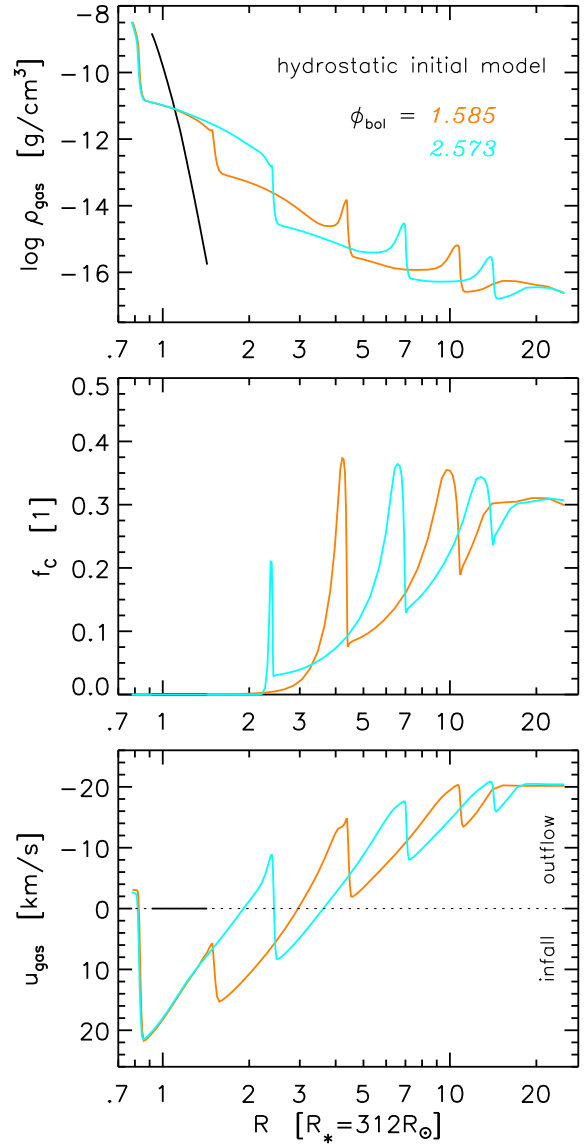}}
\caption{Radial atmospheric structures for the dynamic model used for Fig.\,\ref{f:doubleshellphot} and discussed in App.\,\ref{s:double}. Shown are two characeristic phases (marked with dotted lines in Fig.\,\ref{f:doubleshellphot}) where dust formation takes place ($\phi_{\rm bol}$\,=\,2.573) or not ($\phi_{\rm bol}$\,=\,1.585).}
\label{f:doublestructures}
\end{figure}

In Sect.\,\ref{s:taudust} we illustrated how the time-dependent dust formation in the outer layers of a C-type Mira effects the resulting lightcurves of such an object. This was based on a dynamic model atmosphere with relatively moderate dust formation processes. Here we show a different model\footnote{$L_\star$\,=\,7079\,$L_{\odot}$, $M_\star$\,=\,1\,$M_{\odot}$, $T_\star$\,=\,3000\,K, C/O\,=\,1.69, $P$\,=\,390\,d, $\Delta u_{\rm p}$\,=\,6\,km/s, $f_{\rm L}$\,=\,1}  adopted from the recent grid of Mattsson et al. (\cite{MatWH10}) which represents the quite rare group of strictly `double-periodic' models (cf. the discussion in H\"ofner et al. \cite{HoGAJ03}). As it can be seen in the upper panel of Fig.\,\ref{f:doubleshellphot}, only every second pulsation period a new dust shell emerges and propagates outwards due to the radiation pressure acting upon the dust grains. Figure\,\ref{f:doublestructures} shows that the radial structure of the model replicates in the inner, dust-free layers from cycle to cycle for a given phase $\phi_{\rm bol}$. However, the layers outwards from the dust-forming region at $\approx$\,2--3\,$R_\star$ can differ significantly because of the extreme dust formation. The emergence of the quite narrow dust shell is also reflected in the photometric variations plotted in the lower panel of Fig.\,\ref{f:doubleshellphot} (see also the discussion in Sect.\,\ref{s:taudust}). While the K-lightcurve follows approximately the sinusoidally changing luminosity input, the model exhibits especially in the visual a characteristic lightcurve which is reminiscent of `Miras with secondary maxima' as discussed e.g. by Lebzelter et al. (\cite{LHWJF05a}). It would be interesting to make a more detailed comparison with observed targets showing a similar behaviour, as for example the cases presented in Wood (\cite{WAAAA99}; object 7308.113 in their Fig.\,2), Groenewegen et al. (\cite{Groen04}; object 050709.50-685849.4 in their Fig.\,9), Lebzelter et al. (\cite{LHWJF05a}; R\,Nor in their Fig.\,8), Lebzelter et al. (\cite{LebzW05}; LW10 in their Fig.\,2) or Lebzelter (\cite{Lebze11}; ASAS 201445-4659.0 in their Fig.\,1), in the future. 

\end{appendix}

\listofobjects

\end{document}